# LAW OF LARGE NUMBERS AND CENTRAL LIMIT THEOREM FOR RANDOM SETS OF SOLITONS OF THE FOCUSING NONLINEAR SCHRÖDINGER EQUATION

MANUELA GIROTTI, TAMARA GRAVA, KENNETH D. T-R MCLAUGHLIN, AND JOSEPH NAJNUDEL

Abstract. We study a random configuration of $N$ soliton solutions $\psi_N(x,t;\boldsymbol{\lambda})$ of the cubic focusing Nonlinear Schrödinger (fNLS) equation in one space dimension. The $N$ soliton solutions are parametrized by $2N$ complex numbers $(\boldsymbol{\lambda},\boldsymbol{c})$ where $\boldsymbol{\lambda}\in\mathbb{C}_+^N$ are the eigenvalues of the Zakharov-Shabat linear operator, and $\boldsymbol{c}\in\mathbb{C}^N\backslash\{0\}$ are the norming constants of the corresponding eigenfunctions. The randomness is obtained by choosing the complex eigenvalues to be i.i.d. random variables sampled from a probability distribution with compact support in the complex plane. The corresponding norming constants are interpolated by a smooth function of the eigenvalues. Then we consider the expectation of the random measure associated to this random spectral data. Such expectation uniquely identifies, via the Zakharov-Shabat inverse spectral problem, a solution $\psi_\infty(x,t)$ of the fNLS equation. This solution can be interpreted as a soliton gas solution.

We prove a Law of Large Numbers and a Central Limit Theorem for the differences $\psi_N(x,t;\boldsymbol{\lambda})-\psi_\infty(x,t)$ and $|\psi_N(x,t;\boldsymbol{\lambda})|^2-|\psi_\infty(x,t)|^2$ when $(x,t)$ are in a compact set of $\mathbb{R}\times\mathbb{R}^+$; we additionally compute the correlation functions.

## 1. Introduction

In this manuscript we consider the cubic focusing Nonlinear Schrödinger (fNLS) equation

$$i\psi_t+\frac{1}{2}\psi_{xx}+|\psi|^2\psi=0,\qquad x\in\mathbb{R},\ t\in\mathbb{R}^+, \tag{1.1}$$

with *random soliton initial data* and we establish a Law of Large Numbers and a Central Limit Theorem of its solution for $(x,t)$ in compact sets.

For linear partial differential equations, random initial data is usually constructed from a superposition of uncorrelated linear waves (Fourier modes) with random phases and amplitudes satisfying the Central Limit Theorem. Thanks to the linearity of the differential equation, as time evolves this superposition of linear waves remains uncorrelated and unchanged in distribution.

On the other hand, for nonlinear waves the probability distribution of the wave field deforms substantially in time (see for example the experimental paper [43]). Thus far the evolution has been described for weakly nonlinear waves (i.e. small amplitudes), when the evolution of the expectation of the Fourier modes is described by the wave kinetic equations introduced by Zakharov [50] (see also the books [41], [51]) that have been recently proved for the nonlinear Schrödinger in $d\geq 3$ space dimensions [22].

The nonlinear Schrödinger equation in one space dimension, as with many integrable nonlinear partial differential equations, possesses soliton solutions, and (more interestingly) more complex solutions including multi-soliton solutions, or $N$-soliton solutions, elliptic wave solutions, and dispersive shock waves [8, 9, 14, 38, 39, 40]. These solutions are fundamentally nonlinear, large-amplitude solutions, which exhibit quite complicated behavior (see [5], [6], [7] for solitons and breathers of infinite order). Solutions of the fNLS equation are parametrized via the scattering data (described below), which evolves linearly in time.

A first attempt to analyze solutions to the NLS equation with random initial data can be traced back to the pioneering work of Bourgain [12, 13], where global well-posedness is established for a set of periodic initial data sampled from the (normalized) Gibbs measure. In [30], the authors employ large deviation techniques to analyze the occurrence of rogue waves for the solution of the NLS equation with random periodic initial data in the weakly nonlinear regime.

In this manuscript, we introduce randomness at $t = 0$ through the scattering data, which remains uncorrelated and unchanged in distribution as the solution evolves. This has similarities to the work [18], where a finite Toda lattice with random spectral data was used to study the statistics of deflation times. In a sense, we are introducing randomness in the linear setting of the scattering data, and studying random large-amplitude nonlinear dynamics. An overarching quest is to provide a predictive statistical theory of large amplitude waves in the fNLS equation, over large scales of space and time.

$N$-soliton solutions of integrable nonlinear PDEs have enjoyed a secondary interpretation since the discovery that the KdV equation was integrable in 1967 [29]. In this secondary interpretation, there are $N$ particles, loosely identified with $N$ individual solitons. Integrable techniques have established the following asymptotic behavior for $|t|$ large.

When $t$ is large, either positive or negative, a $N$-soliton solution decomposes into a collection of $N$ well-separated, localized traveling waves. Each traveling wave evolves with a distinct velocity (in the generic case) and so, if one considers larger and larger values of $t$, the distance between them becomes larger and larger as well. Each localized wave is then identified as a particle, with position $x_j(t)$ determined by some identifiable feature, such as the maximum amplitude of the $j$th localized wave.

For intermediate ($\mathcal{O}(1)$) values of $t$, the solution no longer admits this interpretation, since it is not possible to identify $N$ isolated structures in the solution of the PDE. Physically, this is referred to as the interaction or collision of particles. The effect of this interaction is that the $j$th particle emerges with the same velocity, but its position has been shifted by an explicitly calculable amount from what it would have been if no interactions had taken place.

The interpretation of an $N$-soliton solution as a collection of particles led Zakharov to propose a kinetic theory for solitons. Although originally formulated for a dilute gas of solitons for the KdV equation [49], the kinetic theory has been extended to the more general case of a dense gas [24] and to soliton gasses for other equations [25, 26, 27], including the fNLS equation.

The two fundamental ingredients in this kinetic theory are (1) a collection of solitons that are so abundant that they can be described in terms of an evolving "space-time density function" $f(z;x,t)$, and (2) a separate, easily identifiable "tracer soliton" whose velocity, $s(z;x,t)$ depends on the spectral parameter $z$, and is assumed to evolve in $t$ due to the interaction with the gas of solitons. In the end, a coupled system of equations emerges, for the tracer velocity and density:

$$f_t + (sf)_x = 0, \tag{1.2}$$

$$s(z) = -2\mathrm{Re}(z) + \frac{1}{2\mathrm{Im}(z)} \iint \log\left|\frac{z-\overline{w}}{z-w}\right|^2 f(w;x,t)\,[s(z)-s(w)]\,\mathrm{d}^2w \tag{1.3}$$

This system of equations represents the kinetic theory of solitons in the case of the fNLS equation. The equations of the form above, namely the conservation law (1.2) plus an integro-differential equation for the velocity field (1.3), have been named Generalized Hydrodynamic (GHD) equations and they have appeared in the statistical mechanics literature of the last decade [15], [42]. In particular for the discrete nonlinear Schrödinger equation they have been derived in [44], (see also [45] for a survey on classical discrete integrable systems). The GHD equations provide a framework for studying the macroscopic dynamics over large distances and long times of systems that have a microscopic integrable and stationary dynamic. So far, however, the kinetic theory and the generalized hydrodynmic equations are qualitative, and there is to date no rigorous derivation for solitons via analysis of solutions of the underlying nonlinear PDE in the presence of randomness. A rigorous derivation of the kinetic equations in the hydrodynamic limit for a discrete toy model for solitons, called the Box-Ball System, can be found in [16]. Furthermore, there are very recent closely related analytical results [31, 32] for deterministic soliton gasses, which provide a rigorous asymptotic proof of validity of the kinetic equations. It is worth mentioning that during the past 5 years, there have been both numerical simulations of $N$-soliton solutions, and experimental results, which provide compelling confirmation of the kinetic theory (see the review articles [1] and [46]).

In essence, what is missing is a rigorous analysis for *random $N$-soliton solutions* to nonlinear dispersive PDEs, which we develop in this manuscript by establishing a Law of Large Numbers (Theorem 2.6) and a Central Limit Theorem (Theorem 2.7) for solutions of the fNLS equation.

## 2. Statement of results

In 1971, Zakharov and Shabat discovered that the fNLS equation is completely integrable [52]. The following pair of operators forms a Lax pair,

$$\begin{aligned} &\partial_x - \mathcal{L}, \qquad \mathcal{L} = -iz\boldsymbol{\sigma}_3 + \boldsymbol{\Psi}, \quad \boldsymbol{\Psi} = \begin{pmatrix} 0 & \psi \\ -\overline{\psi} & 0 \end{pmatrix} \\ &i\partial_t - \mathcal{B}, \qquad \mathcal{B} = z^2\boldsymbol{\sigma}_3 + iz\boldsymbol{\Psi} + \frac{1}{2}\boldsymbol{\sigma}_3(\boldsymbol{\Psi}^2 - \boldsymbol{\Psi}_x) \ , \end{aligned} \tag{2.1}$$

where $\boldsymbol{\sigma}_3 = \begin{pmatrix} 1 & 0 \\ 0 & -1 \end{pmatrix}$ and $\overline{\psi}$ stands for complex conjugate, so that

$$i\mathcal{L}_t - \mathcal{B}_x + [\mathcal{L}, \mathcal{B}] = i\boldsymbol{\Psi}_t + \frac{1}{2}\boldsymbol{\sigma}_3\boldsymbol{\Psi}_{xx} - \boldsymbol{\sigma}_3\boldsymbol{\Psi}^2 = 0 \ ,$$

which is a restatement of (1.1). For potentials $\psi(x,t)$ that are decaying as $|x| \to \infty$, the scattering and inverse scattering theory of the first operator in (2.1) (the Dirac operator) linearizes the fNLS equation.

The spectrum of the operator (2.1) consists of the real $z$ axis where one defines the reflection coefficient $\rho(z)$, and a finite collection of $L^2$-eigenvalues $\{\lambda_1, \dots, \lambda_N\}$ which are (generically) in the upper half-plane $\mathbb{C}_+$, and for each eigenvalue $\lambda_k$ there is an associated normalization constant $c_k \in \mathbb{C} \smallsetminus \{0\}$. The quantities $\mathcal{S} \coloneqq \left\{\rho(z), \{\lambda_k, c_k\}_{k=1}^N\right\}$ are the scattering data for the potential $\psi$.
The scattering data is determined at $t = 0$. As $\psi$ evolves according to the fNLS equation, the scattering data evolves explicitly in $t$, so that the eigenvalues are constants, and

$$\mathcal{S}(t) = \left\{\rho(z)e^{2itz^2}, \{\lambda_k, c_k e^{2it\lambda_k^2}\}_{k=1}^N\right\}. \tag{2.2}$$

A quick look at the above explicit formulas shows that under the direct scattering transformation, the fNLS equation has been linearized.

The inverse problem is to determine $\psi(x,t)$ from the evolved scattering data $\mathcal{S}(t)$. This inverse problem can be formulated as a *Riemann–Hilbert (RH) problem*. See [10] for a detailed explanation.

The problem is to find a $2 \times 2$ matrix valued function $\boldsymbol{X} = \boldsymbol{X}(z; x, t)$ which satisfies the following properties:

1. $\boldsymbol{X}(z) = \boldsymbol{I} + \mathcal{O}\left(z^{-1}\right)$ as $z \to \infty$, where $\boldsymbol{I}$ is the identity matrix,
2. for $z$ real, $\boldsymbol{X}$ possesses continuous boundary values $\boldsymbol{X}_+(z)$ and $\boldsymbol{X}_-(z)$ (from $\mathbb{C}_\pm$, respectively), which satisfy the jump relation

$$\boldsymbol{X}_+(z) = \boldsymbol{X}_-(z) \begin{pmatrix} 1 + |\rho(z)|^2 & -\overline{\rho(z)}e^{-2itz^2 - 2ixz} \\ \rho(z)e^{2itz^2 + 2ixz} & 1 \end{pmatrix}, \tag{2.3}$$

3. $\boldsymbol{X}$ has simple poles at each $\lambda_k$ and $\overline{\lambda_k}$, where $\boldsymbol{X}$ satisfies a residue condition:

$$\operatorname*{res}_{z=\lambda_k} \boldsymbol{X}(z) = \lim_{z \to \lambda_k} \boldsymbol{X}(z) \begin{pmatrix} 0 & 0 \\ c_k e^{2it\lambda_k^2 + 2ix\lambda_k} & 0 \end{pmatrix}, \tag{2.4a}$$

$$\operatorname*{res}_{z=\overline{\lambda_k}} \boldsymbol{X}(z) = \lim_{z \to \overline{\lambda_k}} \boldsymbol{X}(z) \begin{pmatrix} 0 & -\overline{c_k}e^{-2it(\overline{\lambda_k})^2 - 2ix\overline{\lambda_k}} \\ 0 & 0 \end{pmatrix}. \tag{2.4b}$$

4. $\boldsymbol{X}$ satisfies the Schwarz symmetry

$$\overline{\boldsymbol{X}(\overline{z};x,t)} = \boldsymbol{\sigma}_2 \boldsymbol{X}(z;x,t)\boldsymbol{\sigma}_2 \quad \boldsymbol{\sigma}_2 = \begin{pmatrix} 0 & -i \\ i & 0 \end{pmatrix}. \tag{2.5}$$

This RH problem has a well-established existence and uniqueness theory. The potential $\psi(x,t)$ is extracted from $\boldsymbol{X}$ from the large-$z$ asymptotic behaviour:

$$\boldsymbol{X}(z) = \boldsymbol{I} + \frac{1}{2iz}\begin{pmatrix} -\int_x^\infty |\psi(s,t)|^2 \mathrm{d}s & \psi(x,t) \\ \overline{\psi(x,t)} & \int_x^\infty |\psi(s,t)|^2 \mathrm{d}s \end{pmatrix} + \mathcal{O}\left(\frac{1}{z^2}\right), \tag{2.6}$$

as $z \to \infty$.

The RH formulation of the inverse problem has been used to study asymptotic properties of a wide and ever-growing collection of integrable nonlinear partial differential equations, originating in the work of Deift and Zhou [19]. See [21] for an extension to the perturbed defocusing NLS equation, and [10], [23], [34, 35, 36] for the development and application of $\bar{\partial}$-bar techniques to integrable nonlinear PDEs.

2.1. **Random $N$-soliton solutions.** In this manuscript, we will consider $N$-soliton solutions, for which $\rho(z) \equiv 0$. The scattering data then reduces to the $2N$-dimensional space of eigenvalues and norming constants $\mathcal{S}(t) = \{\lambda_k, c_k e^{2it\lambda_k^2 + 2ix\lambda_k}\}_{k=1}^N$. The fact that the reflection coefficient is identically 0 means that the solution $\boldsymbol{X}$ to the RH problem above is meromorphic in $z$, with simple poles at $\lambda_k$ and $\overline{\lambda_k}$ for each $k = 1, \ldots, N$, and residue conditions (2.4a)-(2.4b).

We consider $N$-soliton solutions with random eigenvalues as follows.

- $\boldsymbol{\lambda} = \{\lambda_1, \ldots, \lambda_N\}$ are i.i.d. random variables sampled according to the uniform distribution over a domain $\mathcal{D}_+ \subset \mathbb{C}_+$

$$\begin{aligned} \mathrm{d}P(\lambda_1, \ldots, \lambda_N) &= \prod_{k=1}^N \mathrm{d}\mu(\lambda_k)\,, \\ \mathrm{d}\mu(z) &= \mathbf{1}_{\mathcal{D}_+}(z)\frac{\mathrm{d}^2 z}{m(\mathcal{D}_+)}, \quad \mathrm{d}^2 z = \mathrm{d}x\mathrm{d}y \end{aligned} \tag{2.7}$$

  where $m(\mathcal{D}_+)$ is the Lebesgue measure of the set $\mathcal{D}_+$, and $\mathbf{1}_{\mathcal{D}_+}(z)$ is the characteristic function of the domain $\mathcal{D}_+$;
- there exists an interpolating function $r \in \mathcal{C}^1(\Omega, \mathbb{C})$, where $\Omega \supset \overline{\mathcal{D}_+}$, with $\overline{\mathcal{D}_+}$ the closure of $\mathcal{D}_+$, such that

$$r(\lambda_k) = N c_k\,, \quad k = 1, \ldots, N. \tag{2.8}$$

*Remark* 2.1. The results which we state below also hold for the case where $\{\lambda_k\}_{k=1}^N$ are i.i.d. random variables, sampled according to a distribution of the form

$$\mathrm{d}\mu(z) = \mathbf{1}_{\mathcal{D}_+}(z)\,\phi(z)\mathrm{d}^2 z\,,$$

for some smooth probability density function $\phi$ with support on $\mathcal{D}_+$. The presence of the factor $\phi(z)$ doesn't alter the proofs, nor does it add further generality.

For each randomly sampled scattering data $\{\lambda_k, \frac{1}{N} r(\lambda_k)\}_{k=1}^N$ at $t = 0$, we consider the solution $\boldsymbol{X}$ of the meromorphic RH problem above, which is now *random*. It is essential that we remove the poles, in favor of jump relations on contours in $\mathbb{C}_+$ and $\mathbb{C}_-$, a fundamental move in the asymptotic analysis of RH problems with poles.

So we introduce a smooth, simple contour $\gamma_+$ in $\mathbb{C}_+$, encircling the domain $\mathcal{D}_+$ (and hence encircling the poles $\{\lambda_k\}_{k=1}^N$ for any configuration). The contour is oriented in the counterclockwise direction. We also introduce the Schwarz-reflected contour $\gamma_-$ in $\mathbb{C}_-$, counterclockwise oriented as well. See Figure 1.

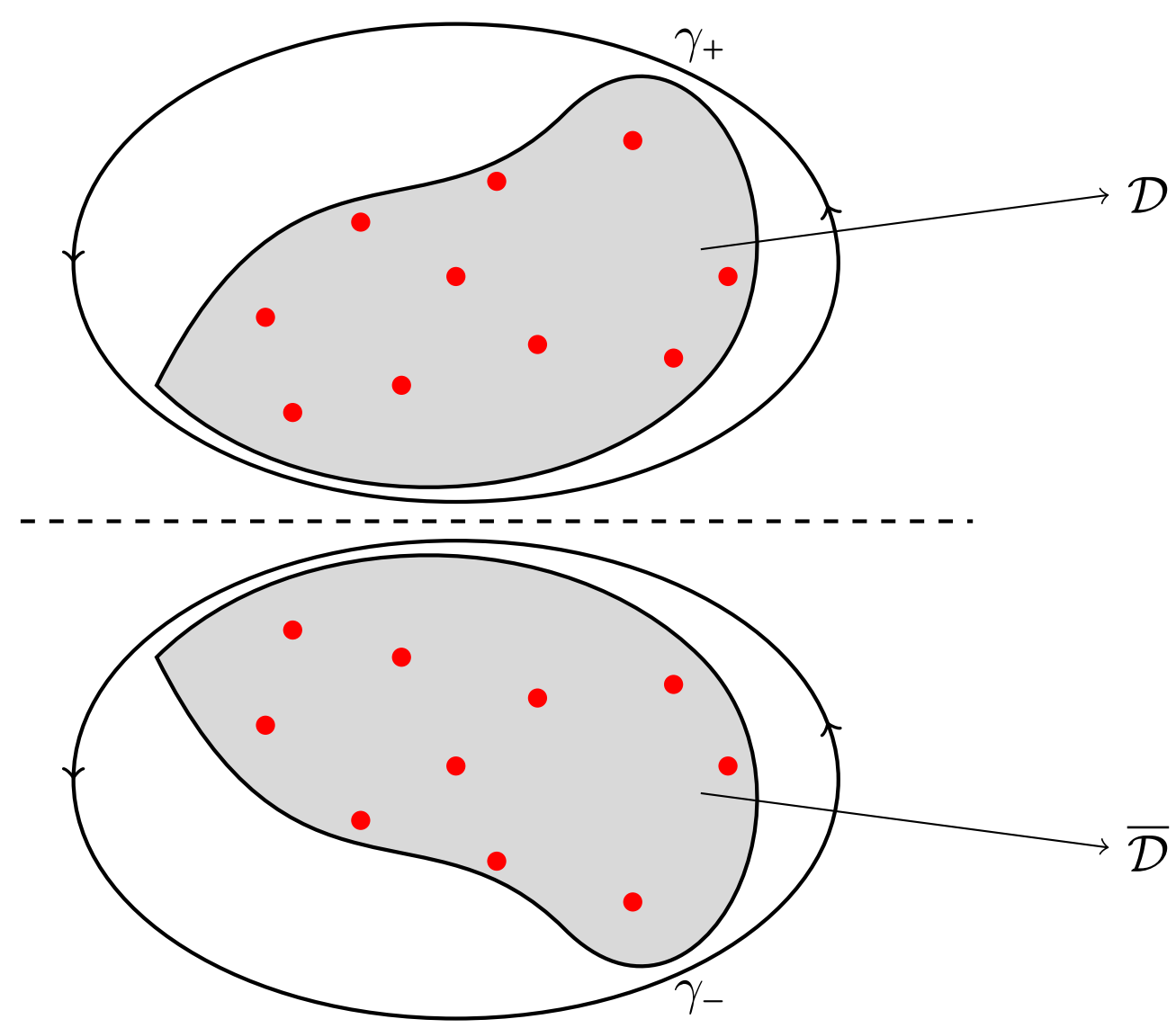

FIGURE 1. The setup: the red poles are sampled uniformly and independently within the region $\mathcal{D}$ (and by symmetry in $\overline{\mathcal{D}}$), as described in (2.7). The contours $\gamma_+$ (resp. $\gamma_-$) encircles $\mathcal{D}$ (resp. $\overline{\mathcal{D}}$) without crossing the real axis.

We also introduce the short-hand notation $\gamma := \gamma_+ \cup \gamma_-$. Instead of $\boldsymbol{X}$, we consider

$$\boldsymbol{M}_N(z) = \boldsymbol{X}(z) \times \begin{cases} \begin{pmatrix} 1 & 0 \\ -e^{\theta(z;x,t)} \sum_{k=1}^{N} \frac{c_k}{z-\lambda_k} & 1 \end{pmatrix} & z \in \operatorname{int}(\gamma_+)\ , \\ \begin{pmatrix} 1 & e^{-\theta(z;x,t)} \sum_{k=1}^{N} \frac{\overline{c_k}}{z-\overline{\lambda_k}} \\ 0 & 1 \end{pmatrix} & z \in \operatorname{int}(\gamma_-)\ , \\ \boldsymbol{I} & \text{otherwise,} \end{cases} \tag{2.9}$$

where

$$\theta(z;x,t) = 2ixz + 2itz^2\,. \tag{2.10}$$

It is straightforward to verify that $\boldsymbol{M}_N(z)$ has no poles in the $z$-plane (the definition is chosen to explicitly cancel each of the poles), and that $\boldsymbol{M}_N(z)$ satisfies the following RH problem.

**Riemann–Hilbert Problem 2.2 (Random $N$-soliton).** Find a $2\times 2$-matrix valued function $\boldsymbol{M}_N = \boldsymbol{M}_N(z;x,t,\boldsymbol{\lambda})$ such that

1. $\boldsymbol{M}_N$ is analytic in $\mathbb{C} \setminus \gamma$.
2. $\boldsymbol{M}_N$ has boundary values $(\boldsymbol{M}_N)_+(z)$ and $(\boldsymbol{M}_N)_-(z)$ for $z$ on the contour $\{\gamma_+ \cup \gamma_-\}$ which satisfy the jump relation

$$(\boldsymbol{M}_N)_+(z) = (\boldsymbol{M}_N)_-(z)\boldsymbol{J}_N(z;x,t,\boldsymbol{\lambda}), \quad z \in \gamma\ . \tag{2.11}$$

with

$$\boldsymbol{J}_N(z;x,t,\boldsymbol{\lambda}) = \begin{cases} \begin{pmatrix} 1 & 0 \\ -e^{\theta(z;x,t)} \sum_{k=1}^{N} \frac{c_k}{z-\lambda_k} & 1 \end{pmatrix}, & z \in \gamma_+ \\ \begin{pmatrix} 1 & e^{-\theta(z;x,t)} \sum_{k=1}^{N} \frac{\overline{c_k}}{z-\overline{\lambda_k}} \\ 0 & 1 \end{pmatrix}, & z \in \gamma_-\,, \end{cases} \tag{2.12}$$

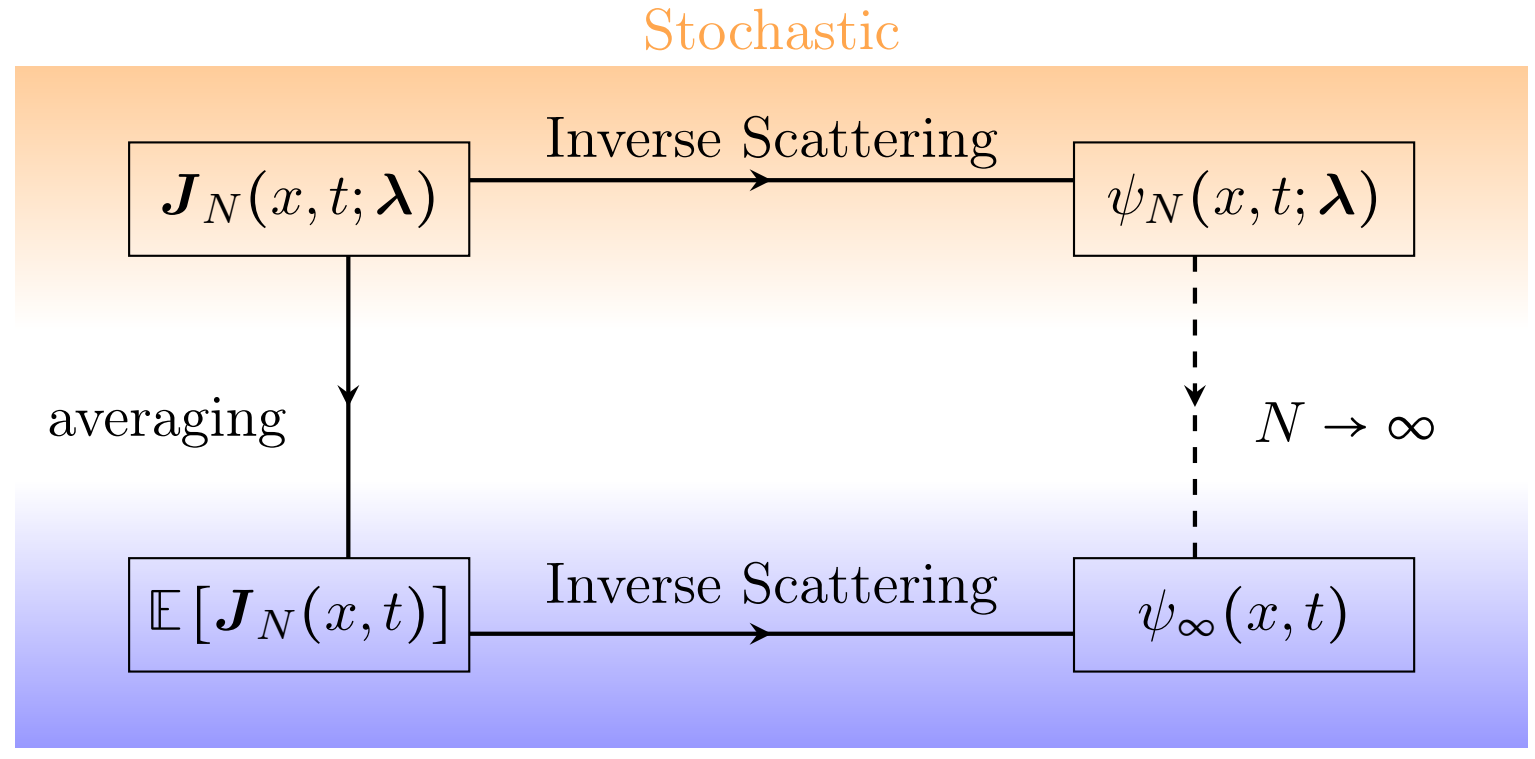

FIGURE 2. A schematic depiction of the setting of Theorem 2.6.

3. $\boldsymbol{M}_N$ satisfies the normalization condition

$$\boldsymbol{M}_N(z;x,t,\boldsymbol{\lambda}) = \boldsymbol{I} + \mathcal{O}\left(\frac{1}{z}\right), \quad \text{as } z \to \infty. \tag{2.13}$$

The $N$-soliton solution $\psi_N(x,t;\boldsymbol{\lambda})$ and its modulus are random variables and they are recovered from the relation

$$\begin{aligned} \psi_N(x,t;\boldsymbol{\lambda}) &= 2i \lim_{z\to\infty} z(\boldsymbol{M}_N(z;x,t,\boldsymbol{\lambda}))_{12}, \\ |\psi_N(x,t;\boldsymbol{\lambda})|^2 &= -2i \lim_{z\to\infty} z\,\partial_x(\boldsymbol{M}_N(z;x,t,\boldsymbol{\lambda}))_{22}, \end{aligned} \tag{2.14}$$

where $(\boldsymbol{M}_N)_{12}$ is the $(1,2)$ entry of the matrix $\boldsymbol{M}_N$, and similarly for the other entries.

We observe however that, due to the nonlinearity of the fNLS equation, the quantity $\mathbb{E}[\psi_N(x,t;\boldsymbol{\lambda})]$ is not a solution of the fNLS equation. Here and below $\mathbb{E}[\cdot]$ stands for the expectation with respect to the probability measure of the eigenvalue distribution.

In order to obtain a deterministic solution to compare to $\psi_N(x,t;\boldsymbol{\lambda})$, we consider a *deterministic inverse scattering problem with the expectation of the jump matrices* in the RH problem. Taking the expectation of the spectral data does not compromise its linear evolution and therefore the solution $\psi_\infty(x,t)$, obtained via inverse scattering, is by construction a solution of the fNLS equation.

We define

$$\boldsymbol{J}(z;x,t) := \mathbb{E}\left[\boldsymbol{J}_N(z;x,t,\boldsymbol{\lambda})\right] , \tag{2.15}$$

and we set up a deterministic RH problem for a matrix $\boldsymbol{M}$ as follows:

**Riemann–Hilbert Problem 2.3 (Averaged RH problem).** Find a $2\times 2$-matrix valued function $\boldsymbol{M} = \boldsymbol{M}(z;x,t)$ such that

1. $\boldsymbol{M}$ is analytic in $\mathbb{C}\setminus\gamma$.
2. $\boldsymbol{M}$ has boundary values $\boldsymbol{M}_+(z)$ and $\boldsymbol{M}_-(z)$ on $\gamma$ which satisfy the jump relation

$$\boldsymbol{M}_+(z) = \boldsymbol{M}_-(z)\boldsymbol{J}(z;x,t), \quad z\in\gamma . \tag{2.16}$$

3. $\boldsymbol{M}$ satisfies the normalization condition

$$\boldsymbol{M}(z) = \boldsymbol{I} + \mathcal{O}\left(\frac{1}{z}\right), \quad \text{as } z\to\infty. \tag{2.17}$$

*Remark* 2.4. Note that, thanks to the eigenvalues $\{\lambda_k\}$'s being i.i.d. and the interpolation (2.8), the averaged jump matrix $\boldsymbol{J}$ does not depend on $N$. Indeed, from the definition (2.15) the jump matrix $\boldsymbol{J}(z;x,t)$ contains terms of the form

$$\mathbb{E}\left[\sum_{k=1}^N \frac{c_k}{z-\lambda_k}\right] = \mathbb{E}\left[\sum_{k=1}^N \frac{r(\lambda_k)}{N(z-\lambda_k)}\right] = \iint_{\mathcal{D}_+} \frac{r(w)}{(z-w)}\mathrm{d}\mu(w) , \tag{2.18}$$

for $z$ outside the closure of $\mathcal{D}_+$. Furthermore, its determinant is identically equal to 1, thanks to the triangular structure of the jump matrices $\boldsymbol{J}_N$, which is preserved after averaging.

**Theorem 2.5** (**Existence of the solution $\psi_\infty$**). *There is a unique solution $\boldsymbol{M}(z;x,t)$ to the Averaged Riemann-Hilbert problem, which determines a solution $\psi_\infty(x,t)$ to the fNLS equation via*

$$\begin{aligned} \psi_\infty(x,t) &= 2i \lim_{z\to\infty} z(\boldsymbol{M}(z;x,t))_{12}\,, \\ |\psi_\infty(x,t)|^2 &= -2i \lim_{z\to\infty} z\,\partial_x(\boldsymbol{M}(z;x,t))_{22}\,. \end{aligned} \tag{2.19}$$

*Moreover, $\psi_\infty$ is a classical solution to the fNLS equation, which belongs to the class $C^\infty(\mathbb{R}\times\mathbb{R}^+)$.*

Existence of the solution $\psi_\infty$ can be proved via an application of the vanishing lemma approach of Zhou [53, Theorem 9.3]. Existence of derivatives of all orders in $x$ and $t$ follows from results proved in much more generality in [20]. In Appendix A, we provide a sketch of the proof.

In a probabilistic sense, the solution $\psi_\infty(x,t)$ can be interpreted as a *soliton gas*, since it coincides with the limit $N\to\infty$ of the $N$-soliton solution. Indeed, the Law of Large Numbers gives

$$\lim_{N\to\infty}\sum_{k=1}^{N}\frac{r(\lambda_k)}{N(z-\lambda_k)} = \iint_{\mathcal{D}_+}\frac{r(w)}{(z-w)}\mathrm{d}\mu(w) \quad \text{almost surely.} \tag{2.20}$$

Thus we can interpret the Averaged RH Problem 2.3 as a gas of solitons whose spectra fill uniformly the domain $\mathcal{D}_+$ (and $\mathcal{D}_-$). The setting is similar to the papers [31, 32], where the authors considered a gas of solitons whose spectra fill in uniformly a segment of the complex plane. In some special cases, the solution $\psi_\infty(x,t)$ can be described quite completely, for example for certain quadrature domains as described in [2, 3].

Our probabilistic results involve comparing $\psi_N(x,t;\boldsymbol{\lambda})$ with $\psi_\infty(x,t)$. See Figure 2.

**Theorem 2.6** (**Convergence in $L^1$**). *Let the eigenvalues $\{\lambda_1,\dots,\lambda_N\}$ of the $N$-soliton solution be sampled according to the probability distribution (2.7) and let the norming constants $\{c_1,\dots,c_N\}$ be interpolated by a $\mathcal{C}^1$ function $r$ according to (2.8). Then the $N$-soliton solution $\psi_N(x,t;\boldsymbol{\lambda})$ and its modulus square $|\psi_N(x,t;\boldsymbol{\lambda})|^2$ converge in mean, as $N\to\infty$, to $\psi_\infty(x,t)$ and $|\psi_\infty(x,t)|^2$ respectively, as defined in (2.19), namely*

$$\lim_{N\to\infty}\mathbb{E}\Big[\,|\psi_N(x,t;\boldsymbol{\lambda})-\psi_\infty(x,t)|\,\Big]=0,$$

*and*

$$\lim_{N\to\infty}\mathbb{E}\Big[|\psi_N(x,t;\boldsymbol{\lambda})|^2-|\psi_\infty(x,t)|^2\Big]=0,$$

*uniformly for $(x,t)$ in a compact set of $\mathbb{R}\times\mathbb{R}^+$.*

Next, we consider the fluctuations of the difference between the random and the deterministic solutions.

**Theorem 2.7** (**Central Limit Theorem**). *Let the points $\{\lambda_1,\dots,\lambda_N\}$ be i.i.d. random variables sampled from the probability distribution (2.7) in the domain $\mathcal{D}_+$, and the norming constants $\{c_k\}$ be interpolated by a $\mathcal{C}^1$ function $r$ according to (2.8). Then the random variables*

$$\sqrt{N}\Big(\psi_N(x,t;\boldsymbol{\lambda})-\psi_\infty(x,t)\Big) \quad \textit{and} \quad \sqrt{N}\Big(|\psi_N(x,t;\boldsymbol{\lambda})|^2-|\psi_\infty(x,t)|^2\Big)$$

*converge in distribution, as $N\to\infty$ and $(x,t)$ in a compact set, to complex and real Gaussian random variables $X^{G_1}$ and $X^{G_2}$ respectively, with zero expectation, and covariance (i.e. expectation of the square)*

$$\mathbb{E}\left[X^{G_i}(x,t)^2\right] = \iint_{\mathcal{D}_+} G_i(w;x,t)^2\mathrm{d}\mu(w) - \left(\iint_{\mathcal{D}_+} G_i(w;x,t)\mathrm{d}\mu(w)\right)^2\,, \tag{2.21}$$

*where the functions $G_1$ and $G_2$ are defined in (5.5) and (5.6). The expectation of the variance (i.e. expectation of the squared modulus) $\mathbb{E}\left[|X^{G_1}(x,t)|^2\right]$ is obtained by replacing the square with the modulus square in the above formula.*

Note that for the real random variable $X^{G_2}$ the definitions of variance and covariance coincide. Note also that the function $G_1$ and $G_2$ defined in (5.5) and (5.6) respectively show the explicit dependence on the solution $\boldsymbol{M}(z;x,t)$ of the averaged RH problem 2.3 and on the interpolating function $r$ in (2.8).

Finally, we calculate the correlation functions.

**Theorem 2.8 (Correlation functions).** *Let $\{\lambda_1,\dots,\lambda_N\}$ be i.i.d. random variables according to the distribution (2.7), and the norming constants $\{c_k\}$ be defined as in (2.8). Let*

$$R_{2,N}(x_1,t_1,x_2,t_2) \coloneqq \mathbb{E}\left[N\left(\psi_N(x_1,t_1;\boldsymbol{\lambda})-\psi_\infty(x_1,t_1)\right)\left(\overline{\psi_N(x_2,t_2)-\psi_\infty(x_2,t_2;\boldsymbol{\lambda})}\right)\right]$$

*be the two-point correlation function. Then, it satisfies*

$$\lim_{N\to\infty} R_{2,N}(x_1,t_1,x_2,t_2) = \iint_{\mathcal{D}_+} G_1(s;x_1,t_1)\overline{G_1(s;x_2,t_2)}\mathrm{d}\mu(s) - \iint_{\mathcal{D}_+} G_1(s;x_1,t_1)\mathrm{d}\mu(s)\overline{\iint_{\mathcal{D}_+} G_1(s';x_2,t_2)\mathrm{d}\mu(s')}. \tag{2.22}$$

*for $(x_i,t_i)$, $i=1,2$, in compact sets of $\mathbb{R}\times\mathbb{R}^+$, where $G_1$ is the complex function defined in (5.5) depending on the solution of the average Riemann-Hilbert problem for $\boldsymbol{M}$ in (2.3).*

*Remark* 2.9. Note that when $(x_1,t_1)=(x_2,t_2)$ one recovers the identity $\lim_{N\to\infty} R_{2,N}(x_1,t_1,x_1,t_1) = \mathbb{E}\left[|X^{G_1}(x_1,t_1)|^2\right]$.

**Outline of the manuscript.** In Section 3 we set up the error problem for the matrix $\boldsymbol{\mathcal{E}}(z) = \boldsymbol{M}_N(z)\boldsymbol{M}(z)^{-1}$ and via a *probabilistic small norm argument* we are able to show the existence of a small norm solution for the RH problem for $\boldsymbol{\mathcal{E}}(z)$ with high probability. This enables the comparison of the two potentials $\psi_N$ and $\psi_\infty$. In Section 4 we prove Theorem 2.6, namely a Law of Large Numbers for the difference $\psi_N(x,t;\boldsymbol{\lambda})-\psi_\infty(x,t)$ and $|\psi_N(x,t;\boldsymbol{\lambda})|^2-|\psi_\infty(x,t)|^2$. In Section 5 we prove Theorem 2.7, namely a Central Limit Theorem for the difference $\sqrt{N}(\psi_N(x,t;\boldsymbol{\lambda})-\psi_\infty(x,t))$ and $\sqrt{N}(|\psi_N(x,t;\boldsymbol{\lambda})|^2-|\psi_\infty(x,t)|^2)$. Finally, in Section 6 we calculate the correlation functions.

## 3. Error analysis and random small norm argument

In order to prove our main results we compare the random RH problem 2.2 for the $N$-soliton solution to the Averaged RH problem 2.3 by considering the error problem

$$\boldsymbol{\mathcal{E}}(z) = \boldsymbol{M}_N(z)\boldsymbol{M}(z)^{-1}. \tag{3.1}$$

This will allow us to directly compare the random potential $\psi_N$ to the deterministic potential $\psi_\infty$ in the limit as $N\to\infty$. Indeed, we have that

$$\psi_N(x,t;\boldsymbol{\lambda})-\psi_\infty(x,t) = 2i\lim_{z\to\infty} z(\boldsymbol{\mathcal{E}}(z;x,t))_{12}, \tag{3.2}$$

$$|\psi_N(x,t;\boldsymbol{\lambda})|^2-|\psi_\infty(x,t)|^2 = 2i\lim_{z\to\infty} z\,\partial_x(\boldsymbol{\mathcal{E}}(z;x,t))_{11} = -2i\lim_{z\to\infty} z\,\partial_x(\boldsymbol{\mathcal{E}}(z;x,t))_{22}, \tag{3.3}$$

namely the knowledge of $\boldsymbol{\mathcal{E}}$ gives information on the difference between the potentials. On the other hand the matrix $\boldsymbol{\mathcal{E}}$ satisfies the following RH problem:

**Riemann–Hilbert Problem 3.1 (Error Problem).** We seek a $2\times2$ matrix-valued function $\boldsymbol{\mathcal{E}} = \boldsymbol{\mathcal{E}}(z;x,t)$ such that

1. $\boldsymbol{\mathcal{E}}$ is analytic in $\mathbb{C}\setminus\gamma$, and it achieves boundary values smoothly on either side of the contours $\gamma_+$ and $\gamma_-$.
2. The boundary values satisfy the jump relation

$$\boldsymbol{\mathcal{E}}_+(z) = \boldsymbol{\mathcal{E}}_-(z)\boldsymbol{J}_{\mathcal{E}}(z;x,t),\ \ z\in\gamma\ , \tag{3.4}$$

$$\boldsymbol{J}_{\mathcal{E}}(z) = \boldsymbol{M}_-(z)\boldsymbol{J}_N(z)\boldsymbol{J}(z)^{-1}\boldsymbol{M}_-(z)^{-1} \tag{3.5}$$

3. $\boldsymbol{\mathcal{E}}$ satisfies the normalization condition

$$\boldsymbol{\mathcal{E}}(z) = \boldsymbol{I} + \mathcal{O}\left(\frac{1}{z}\right) , \quad \text{as } z \to \infty. \tag{3.6}$$

Existence and uniqueness of the matrix $\boldsymbol{\mathcal{E}}$ follows from existence and uniqueness (and invertibility) of the matrices $\boldsymbol{M}_N$ and $\boldsymbol{M}$, by construction. On the other hand, it is desirable to have an explicit estimate of such a solution. To this end, we start by analyzing more in detail the jump matrix $\boldsymbol{J}_{\mathcal{E}}$. We introduce the linear statistics for the function

$$f(w,z) := \frac{r(w)}{z-w}, \tag{3.7}$$

namely

$$X_N^f(z) := \sum_{k=1}^{N} f(\lambda_k, z) - N \iint_{\mathcal{D}_+} f(w,z)\mathrm{d}\mu(w) = \sum_{k=1}^{N} \frac{r(\lambda_k)}{z-\lambda_k} - N \iint_{\mathcal{D}_+} \frac{r(w)}{z-w}\mathrm{d}\mu(w), \tag{3.8}$$

and its Schwarz reflection

$$\overline{X_N^f(\overline{z})} = \sum_{k=1}^{N} \overline{f(\lambda_k, \overline{z})} - N \overline{\iint_{\mathcal{D}_+} f(w,\overline{z})\mathrm{d}\mu(w)} = \sum_{k=1}^{N} \frac{r^*(\lambda_k)}{z-\overline{\lambda_k}} - N \iint_{\mathcal{D}_-} \frac{r^*(w)}{z-w}\mathrm{d}\mu(w) , \tag{3.9}$$

where we used the notation $r^*(w) = \overline{r(\overline{w})}$. Then, the jump matrix $\boldsymbol{J}_{\mathcal{E}}(z)$ takes the compact form

$$\begin{aligned} &\boldsymbol{J}_{\mathcal{E}}(z;x,t) = \boldsymbol{I} + \boldsymbol{W}_N(z;x,t) , \\ &\boldsymbol{W}_N(z) = \frac{1}{N}\boldsymbol{M}_-(z)\begin{pmatrix} 0 & e^{-\theta(z)}\overline{X_N^f(\overline{z})}\mathbf{1}_{\gamma_-}(z) \\ -e^{\theta(z)}X_N^f(z)\mathbf{1}_{\gamma_+}(z) & 0 \end{pmatrix}\boldsymbol{M}_-(z)^{-1} . \end{aligned} \tag{3.10}$$

For simplicity, we will sometimes omit the dependence of $\boldsymbol{W}_N(z;x,t)$ on $x$ and $t$ and write simply $\boldsymbol{W}_N(z)$.

From (3.4) and (3.10), we can express the jump relation for $\boldsymbol{\mathcal{E}}$ as

$$\boldsymbol{\mathcal{E}}_+(z) - \boldsymbol{\mathcal{E}}_-(z) = \boldsymbol{\mathcal{E}}_-(z)\boldsymbol{W}_N(z), \quad z \in \gamma , \tag{3.11}$$

which is equivalently written using the Sokhotski–Plemelj integral formula and the boundary condition (3.6), as follows:

$$\boldsymbol{\mathcal{E}}(z) = \boldsymbol{I} + \frac{1}{2\pi i}\int_{\gamma} \frac{\boldsymbol{\mathcal{E}}_-(s)\boldsymbol{W}_N(s)}{s-z}\mathrm{d}s . \tag{3.12}$$

We can obtain an integral equation by taking the boundary value $\boldsymbol{\mathcal{E}}_-(\xi)$ as $z$ approaches non tangentially the oriented contour $\gamma = \gamma_+ \cup \gamma_-$ from the right:

$$\boldsymbol{\mathcal{E}}_-(\xi) = \boldsymbol{I} + \lim_{\substack{z\to\xi \\ z\in \text{ right side of } \gamma}} \left(\frac{1}{2\pi i}\int_{\gamma} \frac{\boldsymbol{\mathcal{E}}_-(s)\boldsymbol{W}_N(s)}{s-z}\mathrm{d}s\right) . \tag{3.13}$$

By defining the integral operator $c_{\boldsymbol{W}_N}$ as

$$\mathcal{C}_{\boldsymbol{W}_N}(\boldsymbol{h})(\xi) = \mathcal{C}_-(\boldsymbol{h}\boldsymbol{W}_N)(\xi), \tag{3.14}$$

where $\mathcal{C}_-$ is the Cauchy projection operator, namely

$$\mathcal{C}_-(\boldsymbol{h})(\xi) = \lim_{\substack{z\to\xi \\ z\in \text{ right side of } \gamma}} \left(\frac{1}{2\pi i}\int_{\gamma} \frac{\boldsymbol{h}(s)}{s-z}\mathrm{d}s\right) , \tag{3.15}$$

the integral equation (3.13) is then

$$[\mathbf{1} - \mathcal{C}_{\boldsymbol{W}_N}]\boldsymbol{\mathcal{E}}_- = \boldsymbol{I} , \tag{3.16}$$

where $\mathbf{1}$ is the identity operator in $L^2(\gamma)$.

The above expression clearly shows that the existence of a solution $\boldsymbol{\mathcal{E}}_-$ is controlled by the matrix $\boldsymbol{W}_N$, which contains the linear statistic $X_N^f$, and we notice that, when the points $\{\lambda_1,\dots,\lambda_N\}$ are i.i.d. random variables, the Central Limit Theorem [28] guarantees that the random variable $X_N^f(z)/\sqrt{N}$ converges to a complex Gaussian random variable with zero mean and covariance $\mathbb{E}[f(z)^2]-\mathbb{E}[f(z)]^2$.

We will now show that the integral operator in (3.16) is invertible, thus yielding a convergent Neumann series expansion for $\boldsymbol{\mathcal{E}}$ (except for a collection of configurations of $\{\lambda_j\}_{j=1}^N$ whose measure vanishes as $N\to\infty$, see Proposition 4.1). We will resort to a small norm argument [33].

3.1. **Small norm RH theory with high probability.** The goal of this subsection is to show that the matrix $\boldsymbol{W}_N$ defined in (3.10) is small with probability converging to 1 as $N\to\infty$. In this way we can guarantee that the matrix $\boldsymbol{\mathcal{E}}$ can be expressed as a converging Neumann series with probability converging to 1 as $N\to\infty$.

We first consider a uniform estimate for the linear statistic $X_N^f(z)$ of the function $f(w,z)$ defined in (3.7).

Let $\delta>0$. For each fixed $z\in\gamma_+$, let us consider the set

$$B_\delta^\alpha(z)=\left\{\{\lambda_1,\dots,\lambda_N\}:\left|\frac{X_N^f(z)}{N^\alpha}\right|<\delta\right\},\quad 0<\alpha_0<\alpha\le 1, \tag{3.17}$$

where $\alpha_0$ is a fixed number, and define the set

$$B_\delta^\alpha=\bigcap_{z\in\gamma_+}B_\delta^\alpha(z). \tag{3.18}$$

For a configuration of points in $B_\delta^\alpha$, the Schwarz reflection of $X_N^f(z)$ also satisfies the same inequality:

$$\left|(X_N^f)^*(z)\right|=\left|\overline{X_N^f(\overline{z})}\right|<N^\alpha\delta,\qquad \text{for } z\in\gamma_-, \tag{3.19}$$

so that from now on we will only consider $X_N^f(z)$, defined on $\gamma_+$.

Note that $B_\delta^{\alpha_1}(z)\subseteq B_\delta^{\alpha_2}(z)$, for $\alpha_1\le\alpha_2$. We denote simply by $B_\delta$ the set

$$B_\delta:=B_\delta^{\alpha=1}=\bigcap_{z\in\gamma_+}\left\{\{\lambda_1,\dots,\lambda_N\}:\left|\frac{X_N^f(z)}{N}\right|<\delta\right\}\ . \tag{3.20}$$

Following [53], we define the $L^p(\gamma)$-norm of a matrix-valued function as follows. Given the set $\mathrm{Mat}_{2,2}(\mathbb{C})$ of $2\times 2$-matrices, the inner product is

$$(\boldsymbol{A},\boldsymbol{B}):=\mathrm{Tr}(\boldsymbol{B}^*\boldsymbol{A})\qquad\forall\ \boldsymbol{A},\boldsymbol{B}\in\mathrm{Mat}_{2,2}(\mathbb{C})$$

with corresponding norm

$$|\boldsymbol{A}|=\sqrt{\mathrm{Tr}(\boldsymbol{A}^*\boldsymbol{A})}, \tag{3.21}$$

(sometimes called Frobenius norm). Note that any matrix norm satisfies the triangle inequality and the product inequality

$$|\boldsymbol{A}+\boldsymbol{B}|\le|\boldsymbol{A}|+|\boldsymbol{B}|,\quad |\boldsymbol{A}\boldsymbol{B}|\le|\boldsymbol{A}||\boldsymbol{B}|. \tag{3.22}$$

Then, given a function $\boldsymbol{f}:\gamma\to\mathrm{Mat}_{2,2}(\mathbb{C})$, we define its $L^p(\gamma)$-norm $(1\le p<\infty)$ as

$$\|\boldsymbol{f}\|_{L^p(\gamma)}:=\left(\int_\gamma|\boldsymbol{f}(z)|^p\,|\mathrm{d}z|\right)^{\frac{1}{p}}\ ;$$

in particular, for $p=2$ we have

$$\|\boldsymbol{f}\|_{L^2(\gamma)}:=\left(\int_\gamma|f_{11}(z)|^2+|f_{12}(z)|^2+|f_{21}(z)|^2+|f_{22}(z)|^2\,|\mathrm{d}z|\right)^{\frac{1}{2}}\ .$$

On the other hand, for $p=\infty$, we have

$$\|\boldsymbol{f}\|_{L^\infty(\gamma)} := \sup_{z\in\gamma} |\boldsymbol{f}(z)| \ .$$

As shorthand, we may simply indicate $L^p$ instead of $L^p(\gamma)$.

For configurations $\{\lambda_1,\dots,\lambda_N\}$ in the set $B_\delta$ and for $(x,t)$ in a compact set of $\mathbb{R}\times\mathbb{R}^+$ we have

$$\begin{aligned}
\|\boldsymbol{W}_N\|_{L^\infty} &= \sup_{z\in\gamma}\left|\boldsymbol{M}_-(z)\begin{pmatrix} 0 & e^{-\theta(z)}\overline{\frac{X_N^f(\bar z)}{N}}\mathbf{1}_{\gamma_-}(z) \\ -e^{\theta(z)}\frac{X_N^f(z)}{N}\mathbf{1}_{\gamma_+}(z) & 0\end{pmatrix}\boldsymbol{M}_-^{-1}(z)\right| \\
&\le \delta \sup_{z\in\gamma}\left|\boldsymbol{M}_-(z)\begin{pmatrix} 0 & e^{-\theta(z)}\mathbf{1}_{\gamma_-}(z) \\ -e^{\theta(z)}\mathbf{1}_{\gamma_+}(z) & 0\end{pmatrix}\boldsymbol{M}_-^{-1}(z)\right| \\
&\le c_W\delta\ ,
\end{aligned} \tag{3.23}$$

where $c_W>0$ is an absolute constant independent of $N$ and $\delta$. In the above estimate we have used the fact that the second column of $\boldsymbol{M}$ is analytic in $\mathbb{C}_+$ and the first column of $\boldsymbol{M}$ is analytic in $\mathbb{C}_-$. In a similar way using the product matrix norm inequality (3.22) we obtain the estimate

$$\begin{aligned}
&\|\boldsymbol{W}_N\|^2_{L^2(\gamma)} \\
&\le \int_\gamma\left|\boldsymbol{M}_-(z)\begin{pmatrix} 0 & e^{-\theta(z)}\overline{\frac{X_N^f(\bar z)}{N}}\mathbf{1}_{\gamma_-}(z) \\ -e^{\theta(z)}\frac{X_N^f(z)}{N}\mathbf{1}_{\gamma_+}(z) & 0\end{pmatrix}\boldsymbol{M}_-^{-1}(z)\right|^2|\mathrm{d}z| \\
&\le (\widetilde{c_W}\delta)^2\ ,
\end{aligned} \tag{3.24}$$

for some constant $\widetilde{c_W}$ independent from $N$. With the estimate (3.23) we can formulate the following lemma.

**Lemma 3.2.** *Let $\delta>0$. For $(x,t)$ in a compact set of $\mathbb{R}\times\mathbb{R}^+$, there is a constant $c_0$ independent of $N$ and $\delta$ (dependent on the contour $\gamma$ and the function $r$), such that for configurations of points $\{\lambda_1,\dots,\lambda_N\}$ in the set $B_\delta$ defined in (3.20), the Cauchy operator $\mathcal{C}_{\boldsymbol{W}_N}$ defined in (3.14) from $L^2(\gamma)$ to itself has the following uniform bound on the operator norm:*

$$\|\mathcal{C}_{\boldsymbol{W}_N}\|\le c_0\delta. \tag{3.25}$$

*For $c_0\delta<1$ the matrix $\boldsymbol{\mathcal{E}}_-$ is defined by the convergent Neumann series*

$$\boldsymbol{\mathcal{E}}_- = \left(\mathbf{1}-\mathcal{C}_{\boldsymbol{W}_N}\right)^{-1}(\boldsymbol{I}) = \sum_{j=0}^{\infty}\mathcal{C}^j_{\boldsymbol{W}_N}(\boldsymbol{I})\,. \tag{3.26}$$

*Proof.* Let $\{\lambda_1,\dots,\lambda_N\}\in B_\delta$. The Cauchy projection operator $\mathcal{C}_-$ is bounded from $L^2(\gamma)$ to itself by a constant $\mathfrak{C}$ depending only the contour $\gamma$, (see e.g. [11], [17]):

$$\begin{aligned}
\|\mathcal{C}_{\boldsymbol{W}_N}(\boldsymbol{h})\|_{L^2} &= \|\mathcal{C}_-(\boldsymbol{h}\boldsymbol{W}_N)\|_{L^2}\le\mathfrak{C}\|\boldsymbol{h}\boldsymbol{W}_N\|_{L^2} \\
&\le \mathfrak{C}\|\boldsymbol{W}_N(z)\|_{L^\infty}\|\boldsymbol{h}\|_{L^2}\le\mathfrak{C}\left(\delta c_W\right)\|\boldsymbol{h}\|_{L^2}
\end{aligned} \tag{3.27}$$

where $c_W$ has been defined in (3.23). By setting $c_0=\mathfrak{C}c_W$ we have the first statement of the lemma. Next,

$$\|\boldsymbol{\mathcal{E}}_-\|_{L^2} = \left\|\sum_{j=0}^{\infty}\mathcal{C}^j_{\boldsymbol{W}_N}(\boldsymbol{I})\right\|_{L^2}\le(2L_\gamma)^{1/2}\sum_{j=0}^{\infty}(c_0\delta)^j<+\infty \tag{3.28}$$

that is convergent provided that $c_0\delta<1$, where $L_\gamma$ is the length of $\gamma$ and the factor $(2L_\gamma)^{1/2}$ is the $L^2(\gamma)$-norm of the identity matrix. □

From (3.12) and (3.26), we see that, if the configuration is in $B_\delta$ with $c_0\delta < 1$, $\boldsymbol{\mathcal{E}}(z)$ is given by

$$\begin{aligned}\boldsymbol{\mathcal{E}}(z) &= \boldsymbol{I} + \frac{1}{2\pi i}\int_\gamma \frac{\left(\sum_{j=0}^\infty \mathcal{C}^j_{\boldsymbol{w}_N}(\boldsymbol{I})\right)(s)\boldsymbol{W}_N(s)}{s-z}\mathrm{d}s \\ &= \boldsymbol{I} + \int_\gamma \frac{\boldsymbol{W}_N(s)}{s-z}\frac{\mathrm{d}s}{2\pi i} + \int_\gamma \left(\sum_{j=1}^\infty \mathcal{C}^j_{\boldsymbol{w}_N}(\boldsymbol{I})(s)\right)\frac{\boldsymbol{W}_N(s)}{s-z}\frac{\mathrm{d}s}{2\pi i}.\end{aligned} \tag{3.29}$$

From (3.2) and (3.3), we will be interested in the expansion of $\boldsymbol{\mathcal{E}}$ for $z \to \infty$, namely $\boldsymbol{\mathcal{E}}(z) = \boldsymbol{I} + \frac{\boldsymbol{\mathcal{E}}^{(1)}}{z} + O(z^{-2})$. In particular, the $\frac{1}{z}$-term is given by

$$\boldsymbol{\mathcal{E}}^{(1)}(x,t) = -\int_\gamma \boldsymbol{W}_N(s)\frac{\mathrm{d}s}{2\pi i} - \int_\gamma \left(\sum_{j=0}^\infty \mathcal{C}^j_{\boldsymbol{w}_N}(\mathcal{C}_{\boldsymbol{W}_N}(\boldsymbol{I}))\right)\boldsymbol{W}_N(s)\frac{\mathrm{d}s}{2\pi i} \tag{3.30}$$

We are now ready to estimate the difference between $\psi_N(x,t;\boldsymbol{\lambda})$ and $\psi_\infty(x,t)$.

**Proposition 3.3.** *Let $(x,t)$ be in a compact set of $\mathbb{R}\times\mathbb{R}^+$. For all $\epsilon > 0$, there exists $\delta > 0$, independent of $N$, such that for all configurations of random points $\boldsymbol{\lambda} = \{\lambda_1,\dots,\lambda_N\} \in B_\delta$ (with $B_\delta$ the set defined in (3.20)), we have*

$$|\psi_N(x,t;\boldsymbol{\lambda}) - \psi_\infty(x,t)| < \epsilon\ . \tag{3.31}$$

*Proof.* We have

$$\begin{aligned}|\psi_N(x,t;\boldsymbol{\lambda}) - \psi_\infty(x,t)| = |2\boldsymbol{\mathcal{E}}^{(1)}_{12}(x,t;\boldsymbol{\lambda})| &\le \left|\int_\gamma (\boldsymbol{W}_N(s))_{12}\frac{\mathrm{d}s}{\pi}\right| \\ &+ \left|\left(\int_\gamma \left(\sum_{j=0}^\infty \mathcal{C}^j_{\boldsymbol{w}_N}(\mathcal{C}_{\boldsymbol{w}_N}(\boldsymbol{I}))\right)\boldsymbol{W}_N(s)\frac{\mathrm{d}s}{\pi}\right)_{12}\right|.\end{aligned} \tag{3.32}$$

The first term can be easily bounded by

$$\left|\int_\gamma (\boldsymbol{W}_N(s))_{12}\frac{\mathrm{d}s}{\pi}\right| \le \frac{L_\gamma^{1/2}}{\pi}\|\boldsymbol{W}_N\|_{L^\infty}\ . \tag{3.33}$$

Next, we assume $\delta < \frac{1}{c_0\left((2L_\gamma)^{1/2}+1\right)}$, and we use (3.23) , (3.27), and the convergence result (3.28) to obtain

$$\begin{aligned}\left\|\sum_{j=1}^\infty \mathcal{C}^j_{\boldsymbol{w}_N}(\boldsymbol{I})\right\|_{L^2} &= \left\|\mathcal{C}_{\boldsymbol{w}_N}\Big(\sum_{j=0}^\infty \mathcal{C}^j_{\boldsymbol{w}_N}(\boldsymbol{I})\Big)\right\|_{L^2} \\ &\le \mathfrak{C}\,\|\boldsymbol{W}_N\|_{L^\infty}\left\|\sum_{j=0}^\infty \mathcal{C}^j_{\boldsymbol{w}_N}(\boldsymbol{I})\right\|_{L^2} \le (2L_\gamma)^{1/2}\frac{c_0\delta}{1-c_0\delta} < 1\ .\end{aligned} \tag{3.34}$$

Then we estimate the second term in (3.32):

$$\begin{aligned}&\left|\left(\int_\gamma \left(\sum_{j=1}^\infty \mathcal{C}^j_{\boldsymbol{w}_N}(\boldsymbol{I})\right)\boldsymbol{W}_N(s)\frac{\mathrm{d}s}{\pi}\right)_{12}\right| \\ &\le \frac{1}{\pi}\left\|\left(\sum_{j=1}^\infty \mathcal{C}^j_{\boldsymbol{w}_N}(\boldsymbol{I})\right)\boldsymbol{W}_N(s)\right\|_{L^2} \le \frac{\|\boldsymbol{W}_N\|_{L^\infty}}{\pi}\left\|\sum_{j=1}^\infty \mathcal{C}^j_{\boldsymbol{w}_N}(\boldsymbol{I})\right\|_{L^2} \le \frac{\|\boldsymbol{W}_N\|_{L^\infty}}{\pi}\ .\end{aligned} \tag{3.35}$$

We conclude from the above and from (3.23) that

$$|\psi_N(x,t;\boldsymbol{\lambda}) - \psi_\infty(x,t)| = \left|2\boldsymbol{\mathcal{E}}^{(1)}_{12}(x,t)\right| \le \frac{L_\gamma+1}{\pi}\,\|\boldsymbol{W}_N\|_{L^\infty} \le \frac{(L_\gamma^{1/2}+1)c_W\delta}{\pi} \tag{3.36}$$

for configuration $\{\lambda_1, \dots, \lambda_N\}$ in $B_\delta$ and for $(x,t)$ in a compact set of $\mathbb{R} \times \mathbb{R}^+$. It is sufficient to take $\delta < \frac{\pi\epsilon}{(L_\gamma^{1/2}+1)c_W}$ with the constraint $\delta < \frac{1}{((2L_\gamma)^{1/2}+1)c_0}$ to have the statement of the Lemma. □

With little effort we can extend the analysis to the difference $|\psi_N(x,t;\boldsymbol{\lambda})|^2 - |\psi_\infty(x,t)|^2$.

**Lemma 3.4.** *In the same hypotheses as in Lemma 3.2, the solution $\boldsymbol{\mathcal{E}}$ to the RH problem 3.1 is differentiable with respect to $x$ and it admits an expansion in terms of a convergent Neumann series*

$$
\begin{aligned}
\partial_x \boldsymbol{\mathcal{E}}_- &= \left[\mathbf{1} - \mathcal{C}_{\boldsymbol{W}_N}\right]^{-1} \left( \mathcal{C}_{\partial_x \boldsymbol{W}_N} \left( \left[\mathbf{1} - \mathcal{C}_{\boldsymbol{W}_N}\right]^{-1} (\boldsymbol{I}) \right) \right) \\
&= \sum_{j=0}^{\infty} \sum_{k=1}^{j} \mathcal{C}_{\boldsymbol{W}_N}^{k-1} \left( \mathcal{C}_{\partial_x \boldsymbol{W}_N} \mathcal{C}_{\boldsymbol{W}_N}^{j-k} (\boldsymbol{I}) \right) \; .
\end{aligned} \tag{3.37}
$$

*Proof.* The derivative of $\boldsymbol{\mathcal{E}}$ with respect to $x$ needs to satisfy the following RH problem

$$
\begin{aligned}
&\partial_x \boldsymbol{\mathcal{E}}_+(z) = \partial_x \boldsymbol{\mathcal{E}}_-(z) \boldsymbol{J}_{\mathcal{E}}(z;x,t) + \boldsymbol{\mathcal{E}}_-(z) \partial_x \boldsymbol{J}_{\mathcal{E}}(z;x,t), \quad z \in \gamma \; , \\
&\partial_x \boldsymbol{\mathcal{E}}(z) = \mathcal{O}\left(\frac{1}{z}\right) \; , \quad \text{as } z \to \infty.
\end{aligned} \tag{3.38}
$$

This inhomogeneous Riemann-Hilbert problem can be written as a singular integral equation, using the same integral operator as was used for the RH problem 3.1:

$$
\left[\mathbf{1} - \mathcal{C}_{\boldsymbol{W}_N}\right] \partial_x \boldsymbol{\mathcal{E}}_- = \mathcal{C}_{\partial_x \boldsymbol{W}_N} (\boldsymbol{\mathcal{E}}_-) \; , \tag{3.39}
$$

which is invertible in $B_\delta$ via Neumann series, and yields a simple proof of differentiability of $\boldsymbol{\mathcal{E}}$.

Furthermore, similarly as in (3.23) and (3.27), for configurations of points $\{\lambda_1, \dots, \lambda_N\}$ in $B_\delta$ and $(x,t)$ in a compact set of $\mathbb{R} \times \mathbb{R}^+$, we have

$$
\|\partial_x \boldsymbol{W}_N\|_{L^\infty} \leq \widehat{c_W} \delta \; , \qquad \|\mathcal{C}_{\partial_x \boldsymbol{W}_N}\|_{L^2} \leq \widehat{c_0} \delta, \tag{3.40}
$$

for some absolute constants $\widehat{c_W}$ and $\widehat{c_0}$ independent from $\delta$. □

We note in passing that in the appendix we show an alternative route to establishing analyticity in $x$ and $t$ of solutions of RH problems on compact contours, which could equivalently be applied to RH problem 3.1 for configurations in $B_\delta$.

We can now prove an analogue result as Proposition 3.3, but for the difference of the squared modulus of the solutions.

**Proposition 3.5.** *Let $(x,t)$ be in a compact set of $\mathbb{R} \times \mathbb{R}^+$. For all $\epsilon > 0$, there exists $\delta > 0$, independent of $N$, such that for all configurations of random points $\boldsymbol{\lambda} = \{\lambda_1, \dots, \lambda_N\} \in B_\delta$ (with $B_\delta$ the set defined in (3.20)), we have*

$$
\left| |\psi_N(x,t;\boldsymbol{\lambda})|^2 - |\psi_\infty(x,t)|^2 \right| < \epsilon \; . \tag{3.41}
$$

*Proof.* From (3.3) we have

$$
\left| |\psi_N(x,t;\boldsymbol{\lambda})|^2 - |\psi_\infty(x,t)|^2 \right| = 2 \left| \partial_x \boldsymbol{\mathcal{E}}_{22}^{(1)}(x,t;\boldsymbol{\lambda}) \right| \; .
$$

Thanks to Lemma 3.4, we have

$$
\begin{aligned}
\partial_x \boldsymbol{\mathcal{E}}(z) &= \frac{1}{2\pi i} \int_\gamma \frac{\left( \sum_{j=0}^{\infty} \left(\mathcal{C}_{\boldsymbol{W}_N}\right)^j (\boldsymbol{I}) \right)(s) \partial_x \boldsymbol{W}_N(s)}{s - z} \mathrm{d}s \\
&\quad + \int_\gamma \left( \sum_{j=0}^{\infty} \sum_{k=1}^{j} \mathcal{C}_{\boldsymbol{W}_N}^{k-1} \left( \mathcal{C}_{\partial_x \boldsymbol{W}_N} \mathcal{C}_{\boldsymbol{W}_N}^{j-k} (\boldsymbol{I}) \right)(s) \right) \frac{\boldsymbol{W}_N(s)}{s - z} \frac{\mathrm{d}s}{2\pi i}.
\end{aligned} \tag{3.42}
$$

and taking the expansion of $\partial_x \boldsymbol{\mathcal{E}}$ for $z \to \infty$, namely $\partial_x \boldsymbol{\mathcal{E}}(z) = \frac{\partial_x \boldsymbol{\mathcal{E}}^{(1)}}{z} + O(z^{-2})$, the $\frac{1}{z}$ term is given by

$$\begin{aligned}\partial_x \boldsymbol{\mathcal{E}}^{(1)} &= -\int_\gamma \left( \sum_{j=0}^{\infty} \mathcal{C}^j_{\boldsymbol{W}_N}(\boldsymbol{I}) \right)(s) \partial_x \boldsymbol{W}_N(s) \frac{\mathrm{d}s}{2\pi i} \\ &\quad - \int_\gamma \left( \sum_{j=0}^{\infty} \sum_{k=1}^{j} \mathcal{C}^{k-1}_{\boldsymbol{W}_N} \left( \mathcal{C}_{\partial_x \boldsymbol{W}_N} \mathcal{C}^{j-k}_{\boldsymbol{W}_N}(\boldsymbol{I}) \right)(s) \right) \boldsymbol{W}_N(s) \frac{\mathrm{d}s}{2\pi i}.\end{aligned} \tag{3.43}$$

Finally, following closely the steps of Proposition 3.3 and using the estimates (3.40), it is immediate to obtain an $\epsilon$-bound for the difference $|\psi_N(x,t;\boldsymbol{\lambda})|^2 - |\psi_\infty(x,t)|^2$ for configurations in $B_\delta$ with suitable $\delta$. □

Thus far, we have proven that the random solution $\psi_N(x,t;\boldsymbol{\lambda})$ is close to the deterministic solution $\psi_\infty(x,t)$ uniformly for $(x,t)$ in a compact set of $\mathbb{R}\times\mathbb{R}^+$, *provided that the configuration of random points* $\{\lambda_1,\dots,\lambda_N\}$ *is in the set* $B_\delta$.

## 4. Convergence in mean: proof of Theorem 2.6

The goal of this section is to prove convergence in mean of $\psi_N(x,t;\boldsymbol{\lambda})$ and $|\psi_N(x,t;\boldsymbol{\lambda})|^2$, namely

$$\begin{aligned}&\lim_{N\to\infty} \mathbb{E}\Big[ |\psi_N(x,t;\boldsymbol{\lambda}) - \psi_\infty(x,t)| \Big] = 0. \\ &\lim_{N\to\infty} \mathbb{E}\Big[ \left| |\psi_N(x,t;\boldsymbol{\lambda})|^2 - |\psi_\infty(x,t)|^2 \right| \Big] = 0\end{aligned} \tag{4.1}$$

We start by showing that the complement of the set $B^\alpha_\delta$, namely the set

$$(B^\alpha_\delta)^{\mathsf{c}} = \left\{ \{\lambda_1,\dots,\lambda_N\} : \sup_{z\in\gamma_+} \frac{|X^f_N(z)|}{N^\alpha} > \delta \right\}, \tag{4.2}$$

is small as $N \to \infty$. We remind the reader that the linear statistic $X^f_N(z)$ is defined in (3.8), and we observe that now $(B^{\alpha_1}_\delta)^{\mathsf{c}} \supseteq (B^{\alpha_2}_\delta)^{\mathsf{c}}$, when $\alpha_1 \le \alpha_2$.

**Proposition 4.1.** *If the points* $\{\lambda_1,\dots,\lambda_N\}$ *are i.i.d. and distributed according to* (2.7) *then for any* $\delta > 0$ *and integer* $p \ge 1$

$$\begin{aligned}&\mathbb{P}\Big( (B^\alpha_\delta)^{\mathsf{c}} \Big) \le \frac{c_1}{\delta^{2p} N^{p(2\alpha-1)}} + \frac{c_2}{\delta^{2p+1} N^{\alpha(2p+1)-(p+1)}}, \\ &\frac{1}{2} < \frac{p+1}{2p+1} < \alpha \le 1,\end{aligned} \tag{4.3}$$

*for some positive constants* $c_1$ *and* $c_2$ *independent of* $N$ *and* $\delta$ *and depending on the function* $f$ *and the contour* $\gamma_+$.

*Proof.* Given $\delta > 0$ and $\alpha \in (\frac{1}{2}, 1]$ as in (4.3), we define a mesh $\mathcal{M}_M(\gamma_+)$ of $M$ points $\hat{z}_1,\dots,\hat{z}_M$ of the contour $\gamma_+$ so that for all $z \in \gamma_+$, the length of the shortest arc of $\gamma_+$ between $z$ and a point of the mesh is smaller than $\delta/(\tilde{c}N^{1-\alpha})$, for some $\tilde{c}$, independent of $N$ and $\delta$, to be chosen later. It follows that $M$ scales like:

$$M = \mathcal{O}\left( 1 + \frac{L_{\gamma_+} \tilde{c} N^{1-\alpha}}{\delta} \right), \tag{4.4}$$

where $L_{\gamma_+}$ is the length of $\gamma_+$. For any point $z \in \gamma_+$ we have

$$\frac{X^f_N(z)}{N^\alpha} = \frac{X^f_N(\hat{z})}{N^\alpha} + \int_{\hat{z}}^{z} \frac{1}{N^\alpha} \frac{\mathrm{d}X^f_N}{\mathrm{d}w}(w)\,\mathrm{d}w \tag{4.5}$$

where $\hat{z} \in \mathcal{M}_M(\gamma_+)$ is such that the shortest arc between $\hat{z}$ and $z$ has length smaller than $\delta/(\tilde{c}N^{1-\alpha})$, and where the integral from $\hat{z}$ to $z$ is understood as the contour integral on this arc. We get

$$\left|\frac{X_N^f(z)}{N^\alpha}\right| \leq \left|\frac{X_N^f(\hat{z})}{N^\alpha}\right| + \frac{\delta}{\tilde{c}N^{1-\alpha}} \sup_{w\in\gamma_+}\left|\frac{1}{N^\alpha}\frac{\mathrm{d}X_N^f}{\mathrm{d}w}(w)\right| \tag{4.6}$$

where $\frac{\delta}{\tilde{c}N^{1-\alpha}}$ is the upper bound on the arc length between $z$ and $\hat{z}$. Now, from the fact that the distance between points of $\gamma_+$ and points in $\mathcal{D}_+$ is bounded from below, and explicit computation shows that $\frac{1}{N^\alpha}\frac{\mathrm{d}X_N^f}{\mathrm{d}w}(w)$ is dominated by $N$, which gives

$$\sup_{w\in\gamma_+}\left|\frac{1}{N^\alpha}\frac{\mathrm{d}X_N^f}{\mathrm{d}w}(w)\right| < d_0 N^{1-\alpha} \tag{4.7}$$

for some $d_0 > 0$ independent of $N$ and $\delta$. Taking $\tilde{c} = 2d_0$, and assuming to have a configuration $\{\lambda_j\}_{j=1}^N$ in $(B_\delta^\alpha)^\mathsf{c}$, namely $\delta < \frac{|X_N^f(z)|}{N^\alpha}$, we have

$$\delta < \frac{|X_N^f(z)|}{N^\alpha} \leq \frac{|X_N^f(\hat{z})|}{N^\alpha} + \frac{\delta}{2}\,. \tag{4.8}$$

This implies that

$$\frac{\delta}{2} < \frac{|X_N^f(\hat{z})|}{N^\alpha}\,. \tag{4.9}$$

Therefore

$$(B_\delta^\alpha)^\mathsf{c} \subseteq \bigcup_{\hat{z}\in\mathcal{M}_M(\gamma_+)}\left\{\{\lambda_1,\dots,\lambda_N\} : \frac{|X_N^f(\hat{z})|}{N^\alpha} > \frac{\delta}{2}\right\}. \tag{4.10}$$

We now need to estimate $\mathbb{E}[|X_N^f(\hat{z})|^{2p}]$. For points $\{\lambda_1,\dots,\lambda_N\}$ i.i.d. distributed according to (2.7) with interpolating function $r$ as in (2.8), we define the i.i.d. random variables

$$s_k(z) \coloneqq \frac{r(\lambda_k)}{z-\lambda_k} - \iint_{\mathcal{D}_+}\frac{r(w)}{z-w}\mathrm{d}\mu(w)\,, \qquad k = 1,\dots,N\,.$$

Then, the $2p$-th moment is equal to

$$\mathbb{E}\left[|X_N^f(\hat{z})|^{2p}\right] = \mathbb{E}\left[\left|\sum_{k=1}^N s_k(\hat{z})\right|^{2p}\right] = \sum_{k_1,\dots,k_p=1}^N\ \sum_{j_1,\dots,j_p=1}^N \mathbb{E}\left[\prod_{i=1}^p s_{k_i}(\hat{z})\overline{s_{j_i}(\hat{z})}\right] \tag{4.11}$$

it is easy to see that all terms in the sum above are bounded. If an index from $\{k_1,\dots,k_p,j_1,\dots,j_p\}$ is distinct from all the others, the corresponding term vanishes. So the only terms that contribute to the above sum are those for which no index appears exactly once. Now,

$$\#\left\{(k_1,\dots,k_p,j_1,\dots,j_p) : \text{ there are at most } p \text{ distinct indices}\right\} \leq c(p)N^p,$$

where $c(p)$ is independent from $N$. Therefore, since the expectation of each term is uniformly bounded, we conclude that

$$\sup_{\hat{z}_\ell\in\mathcal{M}_M(\gamma_+)}\mathbb{E}\left[|X_N^f(\hat{z}_\ell)|^{2p}\right] \leq c'(p)N^p\,, \tag{4.12}$$

for $c'(p) > 0$ independent of $N$. Finally,

$$\begin{aligned}
\mathbb{P}\Big((B_\delta^\alpha)^{\mathsf{c}}\Big) &\leq \sum_{\ell=1}^{M} \mathbb{P}\left(\left|\frac{X_N(\hat{z}_\ell)}{N^\alpha}\right| > \frac{\delta}{2}\ ,\ \hat{z}_\ell \in \mathcal{M}_M(\gamma_+)\right) \\
&= \sum_{\ell=1}^{M} \mathbb{P}\left(\left|\frac{X_N^f(\hat{z}_\ell)}{N^\alpha}\right|^{2p} > \left(\frac{\delta}{2}\right)^{2p}\ ,\ \hat{z}_\ell \in \mathcal{M}_M(\gamma_+)\right) \\
&\leq \sum_{\ell=1}^{M} 2^{2p} \frac{\mathbb{E}\left[|X_N^f(\hat{z}_\ell)|^{2p}\right]}{\delta^{2p} N^{2p\alpha}} \\
&\leq 2^{2p} N^p \frac{M c'(p)}{\delta^{2p} N^{2p\alpha}}\ ,
\end{aligned} \tag{4.13}$$

where in the third row we have used Markov's inequality, and in the last row we have used (4.12). By substituting $M$ as in (4.4) with $\tilde{c} = 2d_0$ in the above expression, we conclude that (4.3) holds. □

To proceed further we need also a uniform upper bound for the modulus of the $N$-soliton solution $|\psi_N(x,t)|$.

**Lemma 4.2.** *The $N$-soliton solution $\psi_N$ with spectrum $\{\lambda_1,\dots,\lambda_N\}$ satisfies the upper bound*

$$|\psi_N(x,t;\boldsymbol{\lambda})| \leq 4\sum_{k=1}^{N} \operatorname{Im}(\lambda_k) \qquad \forall\,(x,t) \in \mathbb{R}\times\mathbb{R}^+. \tag{4.14}$$

*Proof.* To prove the statement we use the dressing procedure for constructing the $N$-soliton solution with spectrum $\{\lambda_k\}_{k=1}^N$ and the norming constants of the dressing procedure $\{C_k(t)\}_{k=1}^N$, where $C_k(t) = C_k(0)e^{-2i\lambda_k t}$ [37]. The dressing procedure starts from the trivial potential of the fNLS equation, $\psi_{(0)}(x,t) = 0$ for $x \in \mathbb{R}$, and the corresponding matrix solution of the ZS system [52],

$$\boldsymbol{\Phi}^{(0)}(z;x,t) = \begin{pmatrix} e^{-izx} & 0 \\ 0 & e^{izx} \end{pmatrix}; \tag{4.15}$$

At the $n$-th step of the recursive method, the $n$-soliton potential $\psi_n(x,t)$ is constructed via the $(n-1)$-soliton potential $\psi_{n-1}(x,t)$ and the corresponding matrix solution $\boldsymbol{\Phi}^{(n-1)}(z;x,t)$ as

$$\psi_n(x,t) = \psi_{n-1}(x,t) + 2i(\lambda_n - \overline{\lambda_n})\frac{\overline{q_{n1}}\,q_{n2}}{\|\boldsymbol{q}_n\|^2}, \tag{4.16}$$

where the vector $\boldsymbol{q}_n = (q_{n1}, q_{n2})^\top$ is determined by $\boldsymbol{\Phi}^{(n-1)}(z;x,t)$ and the scattering data of the $n$-th soliton $\{\lambda_n, C_n\}$ as

$$\boldsymbol{q}_n(x,t) = \overline{\boldsymbol{\Phi}^{(n-1)}(\overline{\lambda_n};x,t)}\cdot\begin{pmatrix} 1 \\ C_n(t) \end{pmatrix}. \tag{4.17}$$

From the expression (4.16) when $n = N$ we see that

$$\begin{aligned}
|\psi_N(x,t)| &\leq |\psi_{N-1}(x,t)| + 2|\lambda_N - \overline{\lambda_N}|\left|\frac{\overline{q_{n1}}q_{n2}}{\|\boldsymbol{q_n}\|^2}\right| \leq |\psi_{N-1}(x,t)| + 4\operatorname{Im}(\lambda_N) \\
&\leq \sum_{j=1}^{N} 4\operatorname{Im}(\lambda_j).
\end{aligned} \tag{4.18}$$

□

*Remark* 4.3. The above result can be seen as a limiting case of the result in [4] and [48]: finite-gap solutions to the fNLS equation have the modulus bounded by the sum of the imaginary parts of the band endpoints in the upper half plane. In the present case, the bands collapse into single points, hence the extra prefactor. We also mention a similar result for the modulus of the solution to the derivative fNLS equation in [47].

Proposition 3.3 shows that

$$|\psi_N(x,t;\boldsymbol{\lambda}) - \psi_\infty(x,t)| < \epsilon$$

when the configuration of points $\boldsymbol{\lambda} = \{\lambda_1, \dots, \lambda_N\}$ is in the set $B_\delta$. To prove Theorem 2.6, we need to control what happens in the complement of $B_\delta$.

**Proof of Theorem 2.6.** Using Lemma 4.2, Proposition 3.3, and a uniform bound $K_0$ of $|\psi_\infty(x,t)|$ for $(x,t)$ in a given compact set of $\mathbb{R}\times\mathbb{R}^+$, we have that for every $\epsilon > 0$ there is a $\delta > 0$ such that, independently on $N$,

$$\begin{aligned}
\mathbb{E}\Big[\,|\psi_N(x,t;\boldsymbol{\lambda}) - \psi_\infty(x,t)|\,\Big] &= \int_{B_\delta} |\psi_N(x,t;\boldsymbol{\lambda}) - \psi_\infty(x,t)|\,\mathrm{d}P + \\
&\quad + \int_{B_\delta^\complement} |\psi_N(x,t;\boldsymbol{\lambda}) - \psi_\infty(x,t)|\,\mathrm{d}P \\
&\le \epsilon + \int_{B_\delta^\complement} |\psi_N(x,t)|\mathrm{d}P + \int_{B_\delta^\complement} |\psi_\infty(x,t)|\mathrm{d}P \\
&\le \epsilon + (4N \sup_{z\in\mathcal{D}_+} \mathrm{Im}(z) + K_0)\int_{B_\delta^\complement} \mathrm{d}P\,, \qquad (4.19)
\end{aligned}$$

where $P$ is the underlying probability measure. Using the estimates of Proposition 4.1, with $\alpha = 1$ and $p = 2$, we conclude that

$$\mathbb{E}\Big[\,|\psi_N(x,t;\boldsymbol{\lambda}) - \psi_\infty(x,t)|\,\Big] \le \epsilon + (4N \sup_{z\in\mathcal{D}_+} \mathrm{Im}(z) + K_0)\left(\frac{c_1}{\delta^4 N^2} + \frac{c_2}{\delta^5 N^2}\right), \qquad (4.20)$$

for some constants $c_1, c_2$, independent from $N$ and $\delta$. Since $\epsilon$ is arbitrary, we deduce the convergence in mean.

In a similar way, from Proposition 3.5 and Lemma 4.2 we have

$$\begin{aligned}
\mathbb{E}\Big[\,\big|\,|\psi_N(x,t;\boldsymbol{\lambda})|^2 - |\psi_\infty(x,t)|^2\,\big|\,\Big] &= \int_{B_\delta} \Big||\psi_N(x,t;\boldsymbol{\lambda})|^2 - |\psi_\infty(x,t)|^2\Big|\,\mathrm{d}P + \\
&\quad + \int_{B_\delta^\complement} \Big||\psi_N(x,t;\boldsymbol{\lambda})|^2 - |\psi_\infty(x,t)|^2\Big|\,\mathrm{d}P \\
&\le \epsilon + \int_{B_\delta^\complement} |\psi_N(x,t)|^2\mathrm{d}P + \int_{B_\delta^\complement} |\psi_\infty(x,t)|^2\mathrm{d}P \\
&\le \epsilon + \left((4N \sup_{z\in\mathcal{D}_+} \mathrm{Im}(z))^2 + K_0^2\right)\int_{B_\delta^\complement} \mathrm{d}P\,. \qquad (4.21)
\end{aligned}$$

Using the estimates of Proposition 4.1, with $\alpha = 1$ and $p = 3$, we conclude that

$$\mathbb{E}\Big[\,\big|\,|\psi_N(x,t;\boldsymbol{\lambda})|^2 - |\psi_\infty(x,t)|^2\,\big|\,\Big] \le \epsilon + \left((4N \sup_{z\in\mathcal{D}_+} \mathrm{Im}(z))^2 + K_0^2\right)\left(\frac{\widetilde{c_1}}{\delta^6 N^3} + \frac{\widetilde{c_2}}{\delta^7 N^3}\right), \qquad (4.22)$$

for some constants $\widetilde{c_1}, \widetilde{c_2}$, independent from $N$ and $\delta$. □

## 5. Convergence to a Gaussian random variable: proof of Theorem 2.7

We will now show that

$$\sqrt{N}\Big(\psi_N(x,t;\boldsymbol{\lambda}) - \psi_\infty(x,t)\Big)$$

converges to a complex Gaussian random variable with zero mean, and variance and covariance that are explicitly computed function of $(x,t)$, and that

$$\sqrt{N}\Big(|\psi_N(x,t;\boldsymbol{\lambda})|^2 - |\psi_\infty(x,t)|^2\Big)$$

converges to a real Gaussian random variable with zero mean, and explicit variance.

Using (3.2) we obtain

$$\begin{aligned}\sqrt{N}\Big(\psi_N(x,t;\boldsymbol{\lambda})-\psi_\infty(x,t)\Big) &= -\frac{\sqrt{N}}{\pi}\int_\gamma (\boldsymbol{W}_N)_{12}(s;x,t)\mathrm{d}s \\ &\quad -\left[\frac{\sqrt{N}}{\pi}\int_\gamma\left(\sum_{j=0}^{\infty}\mathcal{C}^j_{\boldsymbol{W}_N}(\mathcal{C}_{\boldsymbol{W}_N}(\boldsymbol{I}))\right)\boldsymbol{W}_N(s;x,t)\mathrm{d}s\right]_{12},\end{aligned} \tag{5.1}$$

and

$$\begin{aligned}\sqrt{N}\Big(|\psi_N(x,t;\boldsymbol{\lambda})|^2-|\psi_\infty(x,t)|^2\Big) &= \frac{\sqrt{N}}{\pi}\int_\gamma \partial_x\left(\boldsymbol{W}_N\right)_{22}(s)\mathrm{d}s \\ &+\frac{\sqrt{N}}{\pi}\left[\int_\gamma\left(\sum_{j=1}^{\infty}\mathcal{C}^j_{\boldsymbol{W}_N}(\boldsymbol{I})\right)(s)\partial_x\boldsymbol{W}_N(s)\mathrm{d}s\right]_{22} \\ &+\frac{\sqrt{N}}{\pi}\left[\int_\gamma\left(\sum_{j=0}^{\infty}\sum_{k=1}^{j}\mathcal{C}^{k-1}_{\boldsymbol{W}_N}\left(\mathcal{C}_{\partial_x\boldsymbol{W}_N}\mathcal{C}^{j-k}_{\boldsymbol{W}_N}(\boldsymbol{I})\right)(s)\right)\boldsymbol{W}_N(s)\mathrm{d}s\right]_{22},\end{aligned} \tag{5.2}$$

where the quantity $\boldsymbol{W}_N(z;x,t)$ is defined in (3.10) and we recall that the above Neumann series are convergent in $B_\delta$.

The goal is to show that the first term of the above expressions converges to a Gaussian random variable while the remaining terms become negligible in probability as $N\to\infty$.

Regarding the first term we have the following lemma.

**Lemma 5.1.** *Given the matrix $\boldsymbol{W}_N$ as defined in (3.10), the following identities hold*

$$-\frac{1}{\pi}\int_\gamma(\boldsymbol{W}_N)_{12}(s;x,t)\mathrm{d}s = \frac{1}{N}X_N^{G_1}(x,t) \tag{5.3}$$

$$\frac{1}{\pi}\int_\gamma\partial_x(\boldsymbol{W}_N)_{22}(s;x,t)\mathrm{d}s = \frac{1}{N}X_N^{G_2}(x,t) \tag{5.4}$$

*where $X_N^{G_i}$, $i=1,2$, are the linear statistics*

$$X_N^{G_i}(x,t) := \sum_{j=1}^{N}G_i(\lambda_j;x,t) - N\iint_{\mathcal{D}_+}G_i(w;x,t)\mathrm{d}\mu(w)$$

*of the following functions*

$$G_1(z;x,t) := -2i\left[e^{\theta(z;x,t)}r(z)\boldsymbol{M}_{12}(z;x,t)^2+\overline{e^{\theta(z;x,t)}r(z)\boldsymbol{M}_{22}(z;x,t)^2}\right] \tag{5.5}$$

$$G_2(z;x,t) := -4\partial_x\operatorname{Im}\left[e^{\theta(z;x,t)}r(z)\boldsymbol{M}_{12}(z;x,t)\boldsymbol{M}_{22}(z;x,t)\right]. \tag{5.6}$$

*where $\theta(z)=2ixz+2itz^2$, and the matrix $\boldsymbol{M}(z)$ solves RH problem 2.3.*

*Proof.* We observe that the second column of the matrix $\boldsymbol{M}(z)$ is analytic in $\mathbb{C}_+$ and the first column is analytic in $\mathbb{C}_-$ and furthermore the symmetry (2.5) implies

$$\overline{\boldsymbol{M}_{11}(\overline{z})} = \boldsymbol{M}_{22}(z),\quad \overline{\boldsymbol{M}_{12}(\overline{z})} = -\boldsymbol{M}_{21}(z). \tag{5.7}$$

From the expression (3.10) we have

$$\begin{aligned}(\boldsymbol{W}_N(z))_{12} &= e^{\theta(z)}\frac{X_N^f(z)}{N}\left[\boldsymbol{M}_-(z)\begin{pmatrix}0&0\\-1&0\end{pmatrix}\boldsymbol{M}_-(z)^{-1}\right]_{12} \\ &= e^{\theta(z)}\frac{X_N^f(z)}{N}\boldsymbol{M}_{12}(z)^2,\quad z\in\gamma_+\end{aligned} \tag{5.8}$$

and

$$
\begin{aligned}
(\boldsymbol{W}_N(z))_{12} &= e^{-\theta(z)}\frac{\overline{X_N^f(\overline{z})}}{N}\left[\boldsymbol{M}_-(z)\begin{pmatrix}0 & 1\\ 0 & 0\end{pmatrix}\boldsymbol{M}_-(z)^{-1}\right]_{12}\\
&= e^{-\theta(z)}\frac{\overline{X_N^f(\overline{z})}}{N}\boldsymbol{M}_{11}(z)^2 = e^{-\theta(z)}\frac{\overline{X_N^f(\overline{z})}}{N}\overline{\boldsymbol{M}_{22}(\overline{z})}^2\,,\quad z\in\gamma_-\,. \qquad (5.9)
\end{aligned}
$$

Performing the integral using the residue theorem and the symmetries of $\boldsymbol{M}$ in (5.7) we arrive at

$$
\begin{aligned}
&\frac{1}{2\pi i}\int_\gamma(\boldsymbol{W}_N(s))_{12}\mathrm{d}s =\\
&\frac{1}{N}\sum_{j=1}^N r(\lambda_j)e^{\theta(\lambda_j;x,t)}\boldsymbol{M}_{12}(\lambda_j)^2 - \iint_{\mathcal{D}_+} r(w)e^{\theta(w)}\boldsymbol{M}_{12}(w)^2\mathrm{d}\mu(w)\\
&+\frac{1}{N}\sum_{j=1}^N\overline{r(\lambda_j)e^{\theta(\lambda_j;x,t)}\boldsymbol{M}_{22}(\lambda_j)^2} - \overline{\iint_{\mathcal{D}_+} r(w)e^{\theta(w)}\boldsymbol{M}_{22}(w)^2\mathrm{d}\mu(w)}
\end{aligned}
$$

and the above expression is equivalent with the linear statistic of the function $G_1$ defined in (5.5). In a similar way

$$
(\boldsymbol{W}_N(z))_{22} = e^{\theta(z)}\frac{X_N^f(z)}{N}\boldsymbol{M}_{12}(z)\boldsymbol{M}_{22}(z),\quad z\in\gamma_+
$$

and

$$
(\boldsymbol{W}_N(z))_{22} = e^{-\theta(z)}\frac{\overline{X_N^f(\overline{z})}}{N}\boldsymbol{M}_{11}(z)\boldsymbol{M}_{21}(z)\,,\quad z\in\gamma_-\,.
$$

Performing the integral in (5.4) using the residue theorem and the symmetries of $\boldsymbol{M}$ in (5.7) we arrive at

$$
\begin{aligned}
&\frac{1}{2\pi i}\int_\gamma(\boldsymbol{W}_N(s))_{22}\mathrm{d}s = \Bigg(\frac{1}{N}\sum_{j=1}^N r(\lambda_j)e^{\theta(\lambda_j;x,t)}\boldsymbol{M}_{12}(\lambda_j)\boldsymbol{M}_{22}(\lambda_j)\\
&-\iint_{\mathcal{D}_+} r(w)e^{\theta(w)}\boldsymbol{M}_{12}(w)\boldsymbol{M}_{22}(w)\mathrm{d}\mu(w)\Bigg) - c.c.
\end{aligned}
$$

where *c.c.* stands for complex conjugate. This gives the expression (5.4) . □

**Proof of Theorem 2.7: Central Limit Theorem.** From Lemma 5.1 and equation (5.1) , we can infer that

$$
\begin{aligned}
&\sqrt{N}\Big(\psi_N(x,t;\boldsymbol{\lambda}) - \psi_\infty(x,t)\Big) = \qquad (5.10)\\
&\frac{\sqrt{N}}{N}X_N^{G_1}(x,t) - \left[\frac{\sqrt{N}}{\pi}\int_\gamma\left(\sum_{j=0}^\infty\mathcal{C}_{\boldsymbol{W}_N}^j(\mathcal{C}_{\boldsymbol{W}_N}(\boldsymbol{I}))\right)\boldsymbol{W}_N(s;x,t)\mathrm{d}s\right]_{12}\,.
\end{aligned}
$$

The Central Limit Theorem [28] guarantees that the scaled linear statistic

$$
\frac{1}{\sqrt{N}}X_N^{G_1}(x,t) \qquad (5.11)
$$

converges in distribution to a Gaussian random variable $X^{G_1}$ with zero mean, covariance as in (2.21) and variance (i.e. expectation of the squared modulus)

$$
\mathbb{E}\left[|X^{G_1}(x,t)|^2\right] = \iint_{\mathcal{D}_+}|G_1(w;x,t)|^2\mathrm{d}\mu(w) - \left|\iint_{\mathcal{D}_+}G_1(w;x,t)\mathrm{d}\mu(w)\right|^2. \qquad (5.12)
$$

It remains to prove that the remaining terms in the expansion (5.10) (i.e. the Neumann series) are small in probability. For $\epsilon > 0$, we consider the event

$$F_\epsilon := B_\delta \cap \left\{ \left| \left[ \frac{\sqrt{N}}{\pi} \int_\gamma \left( \sum_{j=0}^{\infty} \mathcal{C}^j_{\boldsymbol{W}_N}(\mathcal{C}_{\boldsymbol{W}_N}(\boldsymbol{I})) \right) \boldsymbol{W}_N(s) \mathrm{d}s \right]_{12} \right| < \epsilon \right\}, \tag{5.13}$$

On the event $F_\epsilon$, we have

$$\sqrt{N}\Big(\psi_N(x,t;\boldsymbol{\lambda}) - \psi_\infty(x,t)\Big) = \frac{1}{\sqrt{N}} X_N^{G_1}(x,t) + \mathcal{O}(\epsilon) \ . \tag{5.14}$$

To conclude the proof, we need to control what happens in the complement of $F_\epsilon$. To this aim, we introduce a $k$-Lipschitz test function $\Phi : \mathbb{C} \to [-1,1]$ (for some number $k > 0$), and we consider the quantity

$$\begin{aligned} &\mathbb{E}\left[\Phi\Big(\sqrt{N}(\psi_N - \psi_\infty)\Big)\right] = \\ &\mathbb{E}\left[\Phi\Big(\sqrt{N}(\psi_N - \psi_\infty)\Big)\mathbf{1}_{F_\epsilon}\right] + \mathbb{E}\left[\Phi\Big(\sqrt{N}(\psi_N - \psi_\infty)\Big)\mathbf{1}_{F_\epsilon^{\mathsf{c}}}\right] \end{aligned} \tag{5.15}$$

(recall that convergence in distribution is implied by the convergence of expectations with respect to arbitrary bounded Lipschitz functions, as in (5.15)).

For the first term in (5.15), since $\Phi$ is $k$-Lipschitz, we have

$$\mathbb{E}\left[\Phi\Big(\sqrt{N}(\psi_N - \psi_\infty)\Big)\mathbf{1}_{F_\epsilon}\right] = \mathbb{E}\left[\left(\Phi\Big(\frac{1}{\sqrt{N}} X_N^{G_1}\Big) + \mathcal{O}(k\epsilon)\right)\mathbf{1}_{F_\epsilon}\right] \ ;$$

for the second term, since $\Phi$ has values in $[-1,1]$, we have

$$\left|\mathbb{E}\left[\Phi\Big(\sqrt{N}(\psi_N - \psi_\infty)\Big)\mathbf{1}_{F_\epsilon^{\mathsf{c}}}\right]\right| \le \mathbb{P}\left(B_\delta^{\mathsf{c}} \cap F_\epsilon^{\mathsf{c}}\right) + \mathbb{P}\left(B_\delta \cap F_\epsilon^{\mathsf{c}}\right) \le \mathbb{P}\left(B_\delta^{\mathsf{c}}\right) + \mathbb{P}\left(B_\delta \cap F_\epsilon^{\mathsf{c}}\right).$$

Then, we deduce

$$\begin{aligned} \mathbb{E}\left[\Phi\Big(\sqrt{N}(\psi_N - \psi_\infty)\Big)\right] &= \mathbb{E}\left[\Phi\left(\frac{1}{\sqrt{N}} X_N^{G_1}\right)\mathbf{1}_{F_\epsilon}\right] + \mathcal{O}\Big(k\epsilon + \mathbb{P}(B_\delta^{\mathsf{c}}) + \mathbb{P}(B_\delta \cap F_\epsilon^{\mathsf{c}})\Big) \\ &= \mathbb{E}\left[\Phi\left(\frac{1}{\sqrt{N}} X_N^{G_1}\right)\right] + \mathcal{O}\Big(k\epsilon + \mathbb{P}(B_\delta^{\mathsf{c}}) + \mathbb{P}(B_\delta \cap F_\epsilon^{\mathsf{c}})\Big) \end{aligned} \tag{5.16}$$

where we have removed the indicator function and absorbed the additional error into $\mathbb{P}\left(B_\delta^{\mathsf{c}}\right) + \mathbb{P}\left(B_\delta \cap F_\epsilon^{\mathsf{c}}\right)$.

Since $\mathbb{P}(B_\delta^{\mathsf{c}})$ tends to zero as $N \to \infty$ thanks to Proposition 4.1, it is enough to prove that for $\epsilon > 0$, $\mathbb{P}(B_\delta \cap F_\epsilon^{\mathsf{c}})$ tends to zero as $N \to \infty$.

For the purpose let us define

$$U(x,t) := \left[ \frac{N}{\pi} \int_\gamma \left( \sum_{j=0}^{\infty} \mathcal{C}^j_{\boldsymbol{W}_N}(\mathcal{C}_{\boldsymbol{W}_N}(\boldsymbol{I})) \right) \boldsymbol{W}_N(s) \mathrm{d}s \ \right]_{12} \ . \tag{5.17}$$

Bounding $|U(x,t)|$ first by the matrix norm and then using the Cauchy-Schwarz inequality for the $L^2(\gamma)$ norm we obtain

$$|U(x,t)| \le \frac{N}{\pi} \int_\gamma \left| \sum_{j=0}^{\infty} \mathcal{C}^j_{\boldsymbol{W}_N}(\mathcal{C}_{\boldsymbol{W}_N}(\boldsymbol{I})) \right| |\boldsymbol{W}_N(s)| |\mathrm{d}s| \le \frac{N}{\pi} \left\| \sum_{j=0}^{\infty} \mathcal{C}^j_{\boldsymbol{W}_N}(\mathcal{C}_{\boldsymbol{W}_N}(\boldsymbol{I})) \right\|_{L^2(\gamma)} \|\boldsymbol{W}_N\|_{L^2(\gamma)} \, .$$

Using the inequality (3.34), the notation of Lemma 3.2 for the norm of the Cauchy operator $c_{\boldsymbol{W}_N}$ and recalling that the norm of the Cauchy projection operator $\mathcal{C}_-$, is $\mathfrak{C}$, we have on $B_\delta$,

$$
\begin{aligned}
|U(x,t)| &\leq \frac{N}{\pi}\sum_{j=0}^{\infty}\left\|\mathcal{C}^j_{\boldsymbol{W}_N}(\mathcal{C}_{\boldsymbol{W}_N}(\boldsymbol{I}))\right\|_{L^2(\gamma)}\|\boldsymbol{W}_N\|_{L^2(\gamma)}\\
&\leq \frac{N}{\pi}\left[\sum_{j=0}^{\infty}(c_0\delta)^j\,\|\mathcal{C}_{\boldsymbol{W}_N}(\boldsymbol{I})\|_{L^2(\gamma)}\right]\|\boldsymbol{W}_N\|_{L^2(\gamma)}\\
&\leq \frac{N}{\pi}\left[\sum_{j=0}^{\infty}(c_0\delta)^j\mathfrak{C}\,\|\boldsymbol{W}_N\boldsymbol{I}\|_{L^2(\gamma)}\right]\|\boldsymbol{W}_N\|_{L^2(\gamma)} = \frac{N\mathfrak{C}}{\pi}\|\boldsymbol{W}_N\|^2_{L^2(\gamma)}\sum_{j=0}^{\infty}(c_0\delta)^j.
\end{aligned}
$$

We chose $\delta$ sufficiently small, namely $c_0\delta < \frac{1}{2}$ so that $\sum_{j=0}^{\infty}(c_0\delta)^j < 2$. Then there is a constant $c > 0$ such that

$$
|U(x,t)| \leq 2\frac{N\mathfrak{C}}{\pi}\|\boldsymbol{W}_N(x,t)\|^2_{L^2(\gamma)} \leq c\sup_{z\in\gamma_+}\frac{|X_N^f(z)|^2}{N}, \tag{5.18}
$$

The constant $c$ is independent from $N$ and $\delta$, so that the above inequality gives a uniform bound of $|U(x,t)|$ for $(x,t)$ in a compact set of $\mathbb{R}\times\mathbb{R}^+$. Then, from Proposition 4.1 (with $\alpha = \frac{3}{4}$ and $p = 4$) we obtain that

$$
\begin{aligned}
\mathbb{P}\left(B_\delta \cap F_\epsilon^c\right) = \mathbb{P}\left(B_\delta, \left|\frac{U(x,t)}{\sqrt{N}}\right| \geq \epsilon\right) \quad &\leq \mathbb{P}\left(c\sup_{z\in\gamma}\frac{|X_N^f(z)|^2}{N^{\frac{3}{2}}} > \epsilon\right)\\
&= \mathbb{P}\left(\sup_{z\in\gamma}\frac{|X_N^f(z)|}{N^{\frac{3}{4}}} > \sqrt{\tfrac{\epsilon}{c}}\right)\\
&\leq \frac{c_1}{N^2\left(\frac{\epsilon}{c}\right)^4} + \frac{c_2}{N^{\frac{7}{4}}\left(\frac{\epsilon}{c}\right)^{\frac{9}{2}}}
\end{aligned} \tag{5.19}
$$

which tends to zero when $N \to \infty$, for any fixed $\epsilon > 0$. Finally, we can conclude that $\sqrt{N}\Big(\psi_N(x,t;\boldsymbol{\lambda}) - \psi_\infty(x,t)\Big)$ converges to a complex Gaussian random variable.

To prove the Central Limit Theorem for the difference of the moduli, from Lemma 5.1, (3.3) and (3.43) we can infer that

$$
\begin{aligned}
&\sqrt{N}(|\psi_N(x,t;\boldsymbol{\lambda})|^2 - |\psi_\infty(x,t)|^2) = -2i\sqrt{N}\partial_x\boldsymbol{\mathcal{E}}^{(1)}_{22}\\
&= \frac{\sqrt{N}}{N}X_N^{G_2}(x,t) + \sqrt{N}\left[\int_\gamma\left(\sum_{j=1}^{\infty}\mathcal{C}^j_{\boldsymbol{W}_N}(\boldsymbol{I})\right)(s)\partial_x\boldsymbol{W}_N(s)\frac{\mathrm{d}s}{\pi}\right]_{22}\\
&+\sqrt{N}\left[\int_\gamma\left(\sum_{j=0}^{\infty}\sum_{k=1}^{j}\mathcal{C}^{k-1}_{\boldsymbol{W}_N}\left(\mathcal{C}_{\partial_x\boldsymbol{W}_N}\mathcal{C}^{j-k}_{\boldsymbol{W}_N}(\boldsymbol{I})\right)(s)\right)\boldsymbol{W}_N(s)\frac{\mathrm{d}s}{\pi}\right]_{22},
\end{aligned} \tag{5.20}
$$

where $X_N^{G_2}(x,t)$ is the linear statistic of the real random variable $G_2$ defined in (5.6). As before, it is a standard fact that $\frac{1}{\sqrt{N}}X_N^{G_2}(x,t)$ converges to a normal distribution $X^{G_2}$ with zero average and variance (2.21). The proof that the remaining terms in the expansion (5.20) (i.e. the Neumann series) are small in probability is similar to the previous case. ◻

## 6. Correlation functions: proof of Theorem 2.8

The final step is the computation of the correlation functions

$$
\begin{aligned}
&R_{2,N}(x_1,t_1,x_2,t_2) =\\
&= \mathbb{E}\Big[N\left(\psi_N(x_1,t_1;\boldsymbol{\lambda}) - \psi_\infty(x_1,t_1)\right)\overline{\left(\psi_N(x_2,t_2;\boldsymbol{\lambda}) - \psi_\infty(x_2,t_2)\right)}\Big].
\end{aligned} \tag{6.1}
$$

We first estimate the correlation function for $\boldsymbol{\lambda} \in B_\delta^{\mathsf{c}}$. From Lemma 4.2 and Proposition 4.1 (with $\alpha = 1$ and $p \geq 4$), we have

$$
\begin{aligned}
&N \left| \int_{B_\delta^{\mathsf{c}}} \left[ \Big(\psi_N(x_1,t_1;\boldsymbol{\lambda}) - \psi_\infty(x_1,t_1)\Big) \overline{\Big(\psi_N(x_2,t_2;\boldsymbol{\lambda}) - \psi_\infty(x_2,t_2)\Big)} \right] \mathrm{d}P \right| \\
&\leq cN^3 \int_{B_\delta^{\mathsf{c}}} \mathrm{d}P \leq cN^3 \left( \frac{c_1}{\delta^{2p} N^p} + \frac{c_2}{\delta^{2p+1} N^p} \right), \quad c > 0
\end{aligned} \tag{6.2}
$$

where the constant c > 0 independent from N follows from the fact that we uniformly bound the $N$-soliton solution by $N$ and $|\psi_\infty(x,t)| \leq K_0$ for some absolute constant $K_0$. Clearly the above quantity goes to zero as $N \to \infty$ with $\delta$ fixed.

Next, we estimate the correlation function in $B_\delta$. By introducing the notation $\xi_j = (x_j, t_j)$, $j = 1, 2$, and using (5.20) and the definition of $U(x,t)$ introduced in (5.17) we have

$$
\begin{aligned}
&N \int_{B_\delta} \Big(\psi_N(\xi_1;\boldsymbol{\lambda}) - \psi_\infty(\xi_1)\Big) \overline{\Big(\psi_N(\xi_2;\boldsymbol{\lambda}) - \psi_\infty(\xi_2)\Big)} \mathrm{d}P = \\
&\frac{N}{\pi^2} \int\limits_{B_\delta} \left( \int\limits_\gamma (\boldsymbol{W}_N)_{12}(s;\xi_1) \mathrm{d}s \overline{\int\limits_\gamma (\boldsymbol{W}_N)_{12}(s';\xi_2) \mathrm{d}s'} \right) \mathrm{d}P + \frac{1}{N} \int\limits_{B_\delta} U(\xi_1) \overline{U(\xi_2)} \mathrm{d}P \\
&\frac{1}{\pi} \int\limits_{B_\delta} \left( \overline{U(\xi_2)} \int\limits_\gamma (\boldsymbol{W}_N)_{12}(s;\xi_1) \mathrm{d}s + U(\xi_1) \overline{\int\limits_\gamma (\boldsymbol{W}_N)_{12}(s;\xi_2) \mathrm{d}s} \right) \mathrm{d}P\,.
\end{aligned} \tag{6.3}
$$

The goal is to show that all terms in the r.h.s of (6.3), except the first one, go to zero as $N \to \infty$. We write $B_\delta$ as the disjoint union of $F_\epsilon$ as given in (5.13) and $B_\delta \cap F_\epsilon^{\mathsf{c}}$. By definition of $F_\epsilon$ in (5.13) and from Lemma 5.1, we have that on $F_\epsilon$ and for $\xi_1$ and $\xi_2$ in a compact set of $\mathbb{R} \times \mathbb{R}^+$,

$$
\begin{aligned}
&\frac{1}{\pi} \int\limits_{F_\epsilon} \left| \overline{U(\xi_2)} \int\limits_\gamma (\boldsymbol{W}_N)_{12}(s;\xi_1) \mathrm{d}s \right| \mathrm{d}P + \frac{1}{\pi} \int\limits_{F_\epsilon} \left| U(\xi_1) \overline{\int\limits_\gamma (\boldsymbol{W}_N)_{12}(s;\xi_2) \mathrm{d}s} \right| \mathrm{d}P \\
&\leq \epsilon \sqrt{N} \int\limits_{F_\epsilon} \left[ \left| \int\limits_\gamma (\boldsymbol{W}_N)_{12}(s;\xi_1) \mathrm{d}s \right| + \left| \int\limits_\gamma (\boldsymbol{W}_N)_{12}(s;\xi_2) \mathrm{d}s \right| \right] \mathrm{d}P \\
&\leq \epsilon\, \mathbb{E} \left[ \left| \frac{X_N^{G_1}(\xi_1)}{\sqrt{N}} \right| \right] + \epsilon\, \mathbb{E} \left[ \left| \frac{X_N^{G_1}(\xi_2)}{\sqrt{N}} \right| \right] \leq c_0 \epsilon
\end{aligned} \tag{6.4}
$$

for some absolute constant $c_0$, since, by the Central Limit Theorem, $X_N^{G_1}/\sqrt{N}$ converges to a Gaussian random variable $X^{G_1}$ with zero mean and variance (5.12). On the event $B_\delta \cap F_\epsilon^{\mathsf{c}}$, we have, using the estimates (3.24), (5.18) and (5.19).

$$
\begin{aligned}
&\frac{1}{\pi} \int\limits_{B_\delta \cap F_\epsilon^{\mathsf{c}}} \left| \overline{U(\xi_2)} \int\limits_\gamma (\boldsymbol{W}_N)_{12}(s;\xi_1) \mathrm{d}s \right| \mathrm{d}P + \frac{1}{\pi} \int\limits_{B_\delta \cap F_\epsilon^{\mathsf{c}}} \left| U(\xi_1) \overline{\int\limits_\gamma (\boldsymbol{W}_N)_{12}(s;\xi_2) \mathrm{d}s} \right| \mathrm{d}P \\
&\leq 4N \mathfrak{C} (\widetilde{c_W} \delta)^3 \int\limits_{B_\delta \cap F_\epsilon^{\mathsf{c}}} \mathrm{d}P \leq \tilde{C} \delta^3 N \frac{c_1}{N^2 \left(\frac{\epsilon}{c_0}\right)^4} + \frac{c_2}{N^{\frac{7}{4}} \left(\frac{\epsilon}{c_0}\right)^{\frac{9}{2}}}
\end{aligned} \tag{6.5}
$$

for some fixed $\epsilon$, $\delta$ and for positive constants $\tilde{C}, c_0, c_1, c_2$ independent from $N$. Clearly the above term goes to zero as $N \to \infty$. Similarly, regarding the second term in the r.h.s. of (6.3), for the configuration in $F_\epsilon$ we have

$$
\frac{1}{N} \int_{F_\epsilon} |U(\xi_1) \overline{U(\xi_2)}| \mathrm{d}P \leq \tilde{c} \epsilon^2\ , \tag{6.6}
$$

for some absolute constant $\tilde{c}$, while for the configuration in $B_\delta \cap F_\epsilon^c$ the integral is bounded by

$$\hat{C}\delta^4 N \int_{B_\delta \cap F_\epsilon^c} \mathrm{d}P \leq \hat{C} N \delta^4 \frac{c_1}{N^2 \left(\frac{\epsilon}{c_0}\right)^4} + \frac{c_2}{N^{\frac{7}{4}} \left(\frac{\epsilon}{c_0}\right)^{\frac{9}{2}}} \,, \tag{6.7}$$

which goes to zero as $N \to \infty$.

We are finally left to evaluate the first term:

$$\begin{aligned}
&\lim_{N\to\infty} \int_{B_\delta} \left[\frac{N}{\pi^2} \left(\int_\gamma (\boldsymbol{W}_N)_{12}(s; x_1, t_1)\mathrm{d}s\right) \left(\overline{\int_\gamma (\boldsymbol{W}_N)_{12}(s'; x_2, t_2)\mathrm{d}s'}\right)\right] \mathrm{d}P \\
&= \lim_{N\to\infty} \int_{B_\delta} \frac{X_N^{G_1}(\xi)}{\sqrt{N}} \frac{\overline{X_N^{G_1}(\xi_2)}}{\sqrt{N}} \mathrm{d}P \\
&= \lim_{N\to\infty} \frac{1}{N} \int_{B_\delta} \left(\sum_{j=1}^N G_1(\lambda_j; \xi_1) - N \iint\limits_{\mathcal{D}_+} G_1(s; \xi_1)\mathrm{d}\mu(s)\right) \times \\
&\qquad\qquad \times \left(\sum_{i=1}^N \overline{G_1(\lambda_i; \xi_2)} - N \overline{\iint\limits_{\mathcal{D}_+} G_1(s; \xi_2)\mathrm{d}\mu(s)}\right) \mathrm{d}P \\
&= \iint\limits_{\mathcal{D}_+} G_1(s; \xi_1)\overline{G_1(s; \xi_2)}\mathrm{d}\mu(s) - \iint\limits_{\mathcal{D}_+} G_1(s; \xi_1)\mathrm{d}\mu(s) \overline{\iint\limits_{\mathcal{D}_+} G_1(s'; \xi_2)\mathrm{d}\mu(s')} \,,
\end{aligned} \tag{6.8}$$

where in the first and second equality we used Lemma 5.1, and in the third equality we use the fact the $\lambda_j$ are i.i.d. random variables. □

*Remark* 6.1. The last line in the formula above can be interpreted as the covariance of the linear Gaussian random variables $X^G(\xi_1)$ and $X^G(\xi_2)$ defined in Theorem 2.7.

## Appendix A. Existence of the solution of the average Riemann-Hilbert problem

**Theorem A.1.** *Given a function $r \in \mathcal{C}^1(\Omega, \mathbb{C})$, $\Omega \supset \overline{\mathcal{D}_+}$, with $\overline{\mathcal{D}_+}$ the closure of $\mathcal{D}_+$, the RH problem 2.3 is uniquely solvable for all $(x,t) \in \mathbb{R} \times \mathbb{R}^+$. Moreover, the function $\psi_\infty(x,t)$ defined in (2.19) is a classical solution to the fNLS equation (1.1), which is actually analytic in both variables.*

*Proof.* The jump matrix $\boldsymbol{J}(z; x, t)$ is analytic for $z \in \gamma_+$, its determinant is identically equal to 1, and the symmetries $\gamma_- = \overline{\gamma_+}$ and $\boldsymbol{J}(z; x, t) = \boldsymbol{J}(\bar{z}; x, t)^\dagger$ are satisfied. Therefore, Zhou's vanishing lemma [53, Theorem 9.3] can be applied to conclude that a unique solution of RH problem 2.3 exists. Uniqueness of the solution follows from a standard Liouville type argument.

Since the jump matrix is analytic, a usual contour deformation argument can be used to show that $\boldsymbol{M}$ is smooth (actually analytic) in $z$ as $z$ approaches the contour $\gamma := \gamma_+ \cup \gamma_-$. Briefly, one considers a slightly deformed contour $\tilde{\gamma}$ which has no intersection with the original contour $\gamma$, where $\boldsymbol{M}$ is uniformly bounded, and then considers the RH problem on $\tilde{\gamma}$, which also possess a unique solution by the same vanishing lemma argument. Because the two different RH problems are related by an explicit and analytic transformation, one learns that the original solution is uniformly bounded in the entire complex plane, and it follows in particular that $\boldsymbol{M}_-$ and its inverse are bounded on $\gamma$.

We next prove analyticity in $x$ and $t$. Let us fix $x_0 \in \mathbb{R}$ and $t_0 \in \mathbb{R}_+$, and consider the ratio

$$\boldsymbol{\mathcal{R}}(z; x, t) = \boldsymbol{M}(z; x, t)\boldsymbol{M}(z; x_0, t_0)^{-1}. \tag{A.1}$$

Since $\boldsymbol{M}(z; x, t)$ and $\boldsymbol{M}(z; x_0, t_0)$ satisfy RH problems on the same contour $\gamma$, the ratio does too, and the jump relation is

$$\begin{aligned}
&\boldsymbol{\mathcal{R}}_+(z) = \boldsymbol{\mathcal{R}}_-(z)\boldsymbol{J}_{\mathcal{R}}(z; x, t, x_0, t_0), \qquad z \in \gamma, \\
&\boldsymbol{J}_{\mathcal{R}}(z; x, t, x_0, t_0) = \boldsymbol{M}_-(z; x_0, t_0)\left[\boldsymbol{J}(z; x, t)\boldsymbol{J}(z; x_0, t_0)^{-1}\right]\boldsymbol{M}_-(z; x_0, t_0)^{-1}.
\end{aligned} \tag{A.2}$$

Now the jump matrix is analytic in the variables $x$ and $t$, and

$$\boldsymbol{J}_{\mathcal{R}}(z; x_0, t_0, x_0, t_0) = \boldsymbol{I} \text{ for all } z \in \gamma. \tag{A.3}$$

Therefore, for all $(x,t)$ close to $(x_0,t_0)$, we know that $\boldsymbol{\mathcal{R}}$ satisfies a small-norm RH problem, which is solvable via Neumann series, and each term in the Neumann series is analytic in $x$ and $t$. It then follows that $\boldsymbol{\mathcal{R}}$ depends analytically on $x$ and $t$, and hence the solution $\psi_\infty(x,t)$ to the nonlinear Schrödinger equation is also analytic in $x$ and $t$.

□

*Remark* A.2. Analyticity of the solution $\boldsymbol{M}$ can be alternatively proven using analytic Fredholm theory.

*Remark* A.3. In [2], the existence of the solution of RH-problem 2.3 was obtained by showing the non vanishing of the $\tau$-function of a $\overline{\partial}$-problem associated to the RH problem 2.3. Such $\tau$-function can be derived as a limit $N \to \infty$ of the $\tau$-function of the $N$-soliton solution, when the soliton spectra is uniformly distributed in the domain $\mathcal{D}_+$.

Furthermore, despite the $\mathcal{C}^\infty$-regularity of the solution $\psi_\infty(x,t)$, the boundary behaviour of the initial profile, i.e. the asymptotic behaviour of $\psi(x,0)$ as $x \to \pm\infty$, remains an open problem. In [2], the large space asymptotics of $\psi_\infty(x,0)$ has been derived for a special choice of $\mathcal{D}_+$.

## Acknowledgments

The authors wish to thank Herbert Spohn for the many insightful discussions on random solitons in integrable equations.
The authors are very grateful to the referee for their thorough review, and for the many valuable suggestions that improved the readability and clarity of the paper.

MG, TG and KM would like to thank the Isaac Newton Institute for Mathematical Sciences, Cambridge, UK and the University of Northumbria, Newcastle, UK for support (EPSRC grant No. EP/V521929/1) and hospitality during the programme "Emergent phenomena in nonlinear dispersive waves" in Summer 2024, where part of the work on this paper was undertaken. MG was partially supported by the Simons Foundation Award SFI-MPS-T-Institutes-00006117 during the programme. MG, TG and KM would also like to thank the American Institute of Mathematics, Pasadena, CA for their hospitality during the SQuaREs program "Integrable PDEs with randomness", where part of their work was done during the 2024 meeting.

MG was supported by the National Science Foundation under Grant No. DMS-2508767. TG acknowledge the support of PRIN 2022 (2022TEB52W) "The charm of integrability: from nonlinear waves to random matrices"-– Next Generation EU grant – PNRR Investimento M.4C.2.1.1 - CUP: G53D23001880006; the GNFM-INDAM group and the research project Mathematical Methods in NonLinear Physics (MMNLP), Gruppo 4-Fisica Teorica of INFN.
Part of this work was completed while KM was visiting the University of Bristol, UK, in Fall 2022 and Fall 2023, and he gratefully thanks the faculty, staff, and administration of the University of Bristol School of Mathematics. Those visits were made possible through the support of a Royal Society Wolfson Fellowship (grant number: `RSWVF\R2\212003`), which KM also gratefully acknowledges. This research was supported in part by the Louisiana Board of Regents Endowed Chairs for Eminent Scholars program.

(Girotti) Department of Mathematics, Emory University, 400 Dowman Dr, Atlanta, GA 30322
*Email address*: `manuela.girotti@emory.edu`

(Grava) SISSA, via Bonomea 265, 34136 Trieste, Italy, INFN sezione di Trieste, and School of Mathematics, University of Bristol, Bristol, BS8 1UG UK
*Email address*: `tamara.grava@bristol.ac.uk`

(McLaughlin) Department of Mathematics, Tulane University, 6823 St Charles Ave, New Orleans, LA 70118
*Email address*: `kmclaughlin@tulane.edu`

(Najnudel) School of Mathematics, University of Bristol, Bristol, BS8 1UG UK
*Email address*: `joseph.najnudel@bristol.ac.uk`